\documentclass{aa}  
\usepackage{graphicx}
\usepackage{color}
\usepackage{graphicx}
\usepackage{lscape}
\usepackage{natbib}
\usepackage[varg]{txfonts}
\usepackage{amssymb}
\usepackage{stmaryrd}
\usepackage{url}
\usepackage{xspace}
\usepackage{color}
\usepackage[usenames,dvipsnames]{xcolor}
\bibpunct{(}{)}{;}{a}{}{,}

\newcommand{\abe}{\object{Abell\,48}\xspace}
\newcommand{\kpd}{\object{KPD\,0005+5106}\xspace}
\newcommand{\ic}{\object{IC\,4663}\xspace}

\newcounter{Rco}
\newcommand{\Ionst}[1]{\setcounter{Rco}{#1}\Roman{Rco}}
\newcommand{\Ion}[2]{\mbox{#1\,{\scriptsize\Ionst{#2}}}}
\newcommand{\Ionw}[3]{\mbox{#1\,{\scriptsize\Ionst{#2}}~$\lambda\,#3$\,\AA}}

\newcommand{\Ionww}[3]{\mbox{#1\,{\scriptsize\Ionst{#2}}~$\lambda\lambda\,#3$\,\AA}}

\newcommand{\logg}{\mbox{$\log g$}\xspace}
\newcommand{\loggw}[1]{\mbox{$\log g\hspace{-0.5mm} =\hspace{-0.5mm}  #1$}}
\newcommand{\kK}{\mbox{\rm kK}}

\newcommand{\se}[1]{\mbox{Sect.\,\ref{#1}}}

\newcommand{\Teff}{\mbox{$T_\mathrm{eff}$}\xspace}
\newcommand{\Teffw}[1]{\mbox{$\Teff\hspace{-0.5mm} =\hspace{-0.5mm} #1 \,\mathrm{kK}$}}

\newcommand{\Msol}{$M_\odot$}

\newcommand{\draft}[1]{
}
\draft{Draft Version: \today.}

\usepackage{amstext}

\begin{document}

\title{Analysis of cool DO-type white dwarfs\\ from the Sloan Digital Sky Survey Data Release 10} 
\author{N\@. Reindl\inst{1}
        \and
        T\@. Rauch\inst{1}
        \and
        K\@. Werner\inst{1}
        \and
        S\@. O\@. Kepler\inst{2}
        \and
        B\@. T\@. G\"ansicke\inst{3}
        \and
        N\@. P\@. Gentile Fusillo\inst{3}}

\institute{Institute for Astronomy and Astrophysics,
           Kepler Center for Astro and Particle Physics,
           Eberhard Karls University, 
           Sand 1,
           72076 T\"ubingen, 
           Germany\\
           \email{reindl@astro.uni-tuebingen.de}
           \and  
           Instituto de F\'{i}sica, Universidade Federal do Rio Grande do Sul, 91501-900 Porto Alegre, RS, Brazil
           \and  
           Department of Physics, University of Warwick, Coventry, CV4 7AL, UK}

\date{Received 26 August 2014; accepted --}

\abstract{We report on the identification of 22 new cool DO-type white dwarfs (WD) detected in Data Release 10 (DR10) of the Sloan Digital Sky Survey 
(SDSS). Among them, we found one more member of the so-called hot-wind DO WDs, which show ultrahigh excitation absorption lines. Our non-LTE model atmosphere analyses 
of these objects and two not previously analyzed hot-wind DO WDs, revealed effective temperatures and gravities in the ranges \Teffw{45-80} and 
\loggw{7.50-8.75}. In eight of the spectra we found traces of C ($0.001-0.01$, by mass). Two of these are the coolest DO WDs  
ever discovered that still show a considerable amount of C in their atmospheres. This is in strong contradiction with diffusion calculations, 
and probably,   similar to what is proposed for DB WDs, a weak mass-loss is present in DO WDs. One object is the most 
massive DO WD discovered so far with a mass of 1.07\Msol\ if it is an ONe-WD or 1.09\Msol\ if it is a CO-WD. We furthermore present the mass distribution of all 
known hot non-DA (pre-) WDs and derive the hot DA to non-DA ratio for the SDSS DR7 spectroscopic sample. The mass distribution 
of DO WDs beyond the wind limit strongly deviates from the mass distribution of the objects before the wind limit. We address this phenomenon by applying different evolutionary input channels. We argue that the DO WD channel may be fed by about 
13\% by post-extreme-horizontal branch stars and that PG\,1159 stars and O(He) stars may contribute in a similar extent to the non-DA WD channel.}


\maketitle 

\section{Introduction}
\label{sect:introduction} 

The vast majority of stars is expected to end as a white dwarf (WD), most of them ($\approx$ 80\%) with 
H-rich atmospheres, corresponding to the DA spectral type. These can be found all along the WD cooling 
sequence, that is, they have $4\,500 \leq \Teff \leq 170\,000$\,K \citep{Sion2011}. In addition, there are the H-deficient WDs 
(non-DA WDs), which are usually divided into three subclasses: the DO spectral type 
($45\,000 \leq \Teff \leq 200\,000$\,K), with the hot DO WDs showing strong \Ion{He}{2} lines, whereas in the spectra of 
cool DO WDs \Ion{He}{1} lines can also be seen; the DB type ($11\,000 \leq \Teff \leq 45\,000$\,K) showing strong 
\Ion{He}{1} lines; and the DC (featureless spectra), DQ, and DZ types ($\Teff \leq 11000$\,K) showing traces of carbon and other metals in 
their spectra \citep{Sion2011}. The overlap in temperature of the hottest DA and non-DA WDs strongly suggests 
the existence of a separate evolutionary channel for both classes. While a direct evolutionary connection for the H-rich 
central stars of planetary nebulae to the DA WDs is very likely \citep{NapiSchoen1995}, the formation and evolution of 
non-DA WDs is less well understood.\\
 DO WDs are commonly believed to be the successors of the PG\,1159 stars (e.g., \citealt{werneretal2014, althausetal2009}), 
which are hot ($75\,000 \leq \Teff \leq 200\,000$\,K) stars that show H-deficient and He-, C-, and O-rich surface 
compositions \citep[typically He\,:\,C\,:\,O = 0.30\,$-$\,0.85\,:\,0.15\,$-$\,0.60\,:\,0.02\,$-$\,0.20 by mass,][]{wernerherwig2006}. 
These abundances can be explained by a very late thermal pulse (VLTP), experienced by a WD during its early cooling 
phase \citep{Iben1983, Althaus2005}. Most of the residual hydrogen envelope is engulfed by the helium-flash 
convection zone and completely burned at the beginning of this thermal pulse. The star is then forced to rapidly evolve 
back to the AGB and finally into a hydrogen-deficient, helium-burning PG\,1159 star \citep{Althaus2005}. As the star 
cools down, gravitational settling removes heavy elements from the photosphere and turns it into a DO WD 
\citep[][unless it is of the subtype hybrid-PG\,1159, then it turns into a DA WD]{UnglaubBues2000}.\\ 
In the past years it became more clear that the non-DA WDs are fed by distinct H-deficient evolutionary channels. In addition to 
the carbon-dominated sequence, a helium-dominated sequence exists \citep{althausetal2009, miszalskietal2012, Reindletal2014, Frewetal2014}. 
VLTP scenarios fail to reproduce the helium-rich atmospheres (He $\geq$ 95\%, by mass) of He-rich sub\-dwarf O (sdO) stars, R Coronae Borealis (RCB) stars, 
extreme helium (EHe) stars, [WN]-type central stars, and O(He) stars, suggesting that these objects have a different formation history. 
The origin of these stars remains uncertain. Their abundances match predictions of a double-helium WD merger 
scenario \citep{zhangetal2012b, zhangetal2012a}, suggesting the evolutionary channel sdO(He) 
$\rightarrow$ O(He) $\rightarrow$ DO WD or, in case of C- and N-rich sdO and O(He) stars, RCB $\rightarrow$ EHe $\rightarrow$ sdO(He) 
$\rightarrow$ O(He) $\rightarrow$ DO WD \citep{Reindletal2014}. The existence of planetary nebulae that do not show helium enrichment around every 
other O(He) star or [WN]-type central star, however, precludes a double-helium WD merger origin for these stars. These stars must 
have formed in a different way, for instance, by enhanced mass-loss during their post-AGB evolution, or a merger within a common
envelope 
of a CO-WD and a red giant or AGB star \citep{Reindletal2014}.\\
\cite{Dufour2007} reported the discovery of several WDs with atmospheres primarily composed of carbon, with little or no trace of 
hydrogen or helium. These stars do not fit satisfactorily in any of the currently known theories of post-AGB evolution. They are 
considered to be the cooler counterpart of the unique and extensively studied PG1159 star \object{H1504+65}\xspace \citep{Nousek1986, Werner1991, Werner2004a} 
and might form a new evolutionary sequence that follows the AGB. Another case of H-deficient WDs that need to fit into the evolutionary picture 
are the oxygen-dominated WDs discovered by \cite{Gaensicke2010}.\\
As a result of their fast evolutionary rate, there are only a few very hot H-deficient stars. A literature study revealed that there are currently only 46 known
PG\,1159 stars, 10 O(He) stars, and 52 DO WDs\footnote{This excludes the 22 new DO WDs from this paper.}$^{,}$\footnote{A list of all analyzed DO WDs can be found 
at \url{http://astro.uni-tuebingen.de/~reindl/He}\,.}. The detection and analysis of new 
DO WDs improve their statistics and thus helps to understand the origin of those objects. Furthermore, it is of importance for the construction of the 
hot end of the WD luminosity function. Its shape is an excellent tool for constraining the emission 
of particles in the core of hot DO WDs, for example, and to check for the possible existence of DFSZ axions, a proposed but not yet detected type of weakly interacting 
particles \citep{MillerBertolami2014a, MillerBertolami2014b}.

In this paper, we first describe the observations and line identifications (\se{sect:observation}). The spectral analysis follows in 
\se{sect:analysis}. The results are discussed in \se{sect:discussion}, where we mention the phenomenon of the so-called hot-wind DO WDs (\se{subsect:uhei}) 
and debate the evolution of the C abundances before and along the non-DA WD cooling sequence (\se{subsect:C}). The mass distribution of all known O(He) stars, 
PG\,1159 stars, and DO WDs is presented in \se{subsect:mass} and the hot DA to non-DA ratio in \se{subsect:ratio}.\\
In the text, we use abbreviated versions of the object names. Full names are given in the Tables\,\ref{tab:nyaDOs} and \ref{tab:parameters}.

\section{Observations and line identifications}
\label{sect:observation} 

Until 1996, only 22 DO WDs\footnote{This excludes \kpd which was recently reclassified as a pre-WD \citep{werneretal2014} and is now considered 
as an O(He) star \citep{Reindletal2014}.} were known \citep{dreizlerwerner1996}. 
\cite{huegelmeyeretal2005, huegelmeyeretal2006} almost doubled this number with objects from the SDSS DR1, DR2, DR3, 
and DR4. Recently, \cite{werneretal2014} detected ten new hot DO WDs in the SDSS DR 10. To complete the sample of DO WDs, 
we have visually scanned a color-selected sample of WD candidates in the SDSS DR10 and also searched for cool DO WDs. 
These are distinguished from the hot WDs by the presence of \Ion{He}{1} lines. To distinguish cool DO WDs from 
sdO stars, we only considered spectra with broad and only weak higher order (n\,$\ge$ 10) \Ion{He}{2} lines. 
We found 24 objects (Figs.~\ref{fig:do1} and 2) that have not previously been analyzed with non-LTE model atmospheres. 
This sample includes two previously not analyzed hot-wind DO WDs (see below) and almost doubles the number of cool 
DO WDs below 80\,kK.\\
Of the 24 stars, 14 were already included in the DR7 and also found by \cite{Kleinman2013}. These authors classified 11 of 
the stars as DO WDs, two as DOBAH (J1442, J1509) and one as DBAH (J0005). The latter was already included in the SDSS DR4 catalog of 
spectroscopically confirmed WDs \citep{Eisenstein2006} and classified as an sdO star. The analyses of \cite{Kleinman2013} and \cite{Eisenstein2006} 
were only based on LTE models and, thus do not provide reliable atmospheric parameters for DO WDs. \cite{Kleinman2013} listed 52 additional DO WDs, 
28 of them previously analyzed with non-LTE model atmospheres. The remaining 24 had spectra with a too low signal-to noise (S/N) ratio or were either 
a misclassified cataclysmic variable (CV), misclassified He-sdO stars (\logg $\leq 7.0$), or DAO WDs. J0839 is listed as DAO WD, but might be a good 
candidate for a DO WD. The poor S/N, however, does not allow a precise spectral analysis. All these objects are listed in Table~\ref{tab:nyaDOs} 
and were rejected from our sample.\\

\begin{table}[ht]\centering
\caption{DO WD candidates from the SDSS DR7 WD catalog \citep{Kleinman2013}
that were rejected from our sample.}
\label{tab:nyaDOs}
\renewcommand{\tabcolsep}{1.5mm}
\begin{tabular}{l c c l} 
SDSS name            & $g$ [mag] & S/N & comment   \\ 
\hline 
\noalign{\smallskip}
\object{J012602.53$-$004834.5}\xspace  & 17.97 & 20  & misclassified He$-$sdO \\
\object{J014531.86+010629.7}\xspace    & 20.15 & \,\,8   & poor quality spectrum \\
\object{J024323.23+275045.5}\xspace    & 19.48 & 12  & poor quality spectrum \\
\object{J025622.18+330944.7}\xspace    & 19.70 & 10   & poor quality spectrum \\
\object{J065745.83+834958.5}\xspace    & 19.18 & 18  & poor quality spectrum \\
\object{J080846.19+313106.0}\xspace   & 19.44 & 53 &  misclassified CV\\
\object{J081533.08+264646.4}\xspace   & 19.48 & 19 &  poor quality spectrum\\
\object{J083959.93+142858.0}\xspace    & 18.61 & \,\,6  & poor quality spectrum \\
\object{J094526.91+172917.2}\xspace   & 20.28 & \,\,5  & poor quality spectrum \\
\object{J102907.31+254008.4}\xspace  &  17.35   & 16     & misclassified DAO \\
\object{J123524.29-011547.4}\xspace  &  19.15   & \,\,9  & poor quality spectrum \\
\object{J130249.00-013309.5}\xspace  &  18.71   & 11     & poor quality spectrum \\
\object{J131816.55+485741.3}\xspace    & 19.15 & \,\,7   & poor quality spectrum \\
\object{J151246.56+071517.3}\xspace  &  19.41   & 13     & poor quality spectrum \\
\object{J154829.87+203139.1}\xspace  &  16.81   & 35     & misclassified DAO \\
\object{J155642.95+501537.5}\xspace    & 15.81 & 43  & misclassified He$-$sdO \\
\object{J161512.22+110240.0}  & 16.89   & 36     & misclassified DAO \\
\object{J163200.32$-$001928.3}\xspace  & 18.37 & 20  & misclassified DAO \\
\object{J171600.52+422131.1}\xspace    & 18.19 & 16  & poor quality spectrum \\
\object{J173027.20+265639.5}\xspace    & 17.07 & 33  & misclassified He$-$sdO \\
\object{J173824.64+581801.8}  & 18.36   & 13     & poor quality spectrum \\
\object{J205030.40$-$061957.9}\xspace  & 17.98 & 21  & misclassified DAO \\
\object{J205930.25$-$052848.9}\xspace  & 17.64 & 25  & misclassified DAO \\
\object{J213932.49+112611.3} &  19.29   &  9     & poor quality spectrum \\ 
\noalign{\smallskip}
\hline
\end{tabular}
\end{table}

Table~\ref{tab:parameters} lists all the DO WDs from our sample. 
In the spectra of six newly discovered objects (J0742, J0902, J1107, J1531, J1707, and J1717), we were able to detect the 
\Ionww{C}{4}{4647, 4657, 4658, 4659, 4660, 5803, 5814} lines. For the first time, we identified \ion{C}{III} lines in the spectra of two DO WDs. 
In the spectrum of J2239, we found \Ionww{C}{3}{4515, 4516, 4517, 4647, \text{and }   4650}, and in the spectrum of J0301, we additionally identified 
\ion{C}{III}\,$\lambda\lambda$ 4056, 4068, 4069, 4070, 4187,
and 4326\,\AA.\\
Our sample furthermore includes three members of the so-called hot-wind DO WDs, which show ultrahigh excitation (uhei) absorption lines. 
The uhei lines in the spectrum of J0747 (\object{HS\,0742+6520}\xspace) were identifed for the first time. 
This star was first classified as an O(He) star by \cite{Heber1996}, who discovered it in the Hamburg-Schmidt survey. 
However, because of the low S/N, the uhei features were not clearly visible in this spectrum. \cite{Sinamyan2011} used empirical formulas 
to estimate \Teff\ and \logg\ for 87 First Byurakan Survey (FBS) WDs. Based on the SDSS colors, he found for J0747 \Teff$=$\,85\,279\,K 
and \loggw{7.78}. J0201 (\object{HS\,0158+2335}\xspace, discovered by \citealt{Dreizleretal1995}) and J0717 (also known as \object{HS\,0713+3958}\xspace, 
discovered by \citealt{werneretal1995}) are known DO WDs with uhei features. Since their SDSS spectra 
range up to 10\,000\,\AA,\, we were able to identify additional lines beyond the hitherto observed H$\alpha$ region (Table~\ref{tab:uhei}).
Figure~\ref{fig:do3} shows the spectra of these DO WDs.

\begin{landscape}
\addtolength{\textwidth}{6.3cm} 
\addtolength{\evensidemargin}{-3cm}
\addtolength{\oddsidemargin}{-3cm}
\begin{figure*}
\includegraphics[trim=0 0 0 -120,height=20.0cm,angle=0]{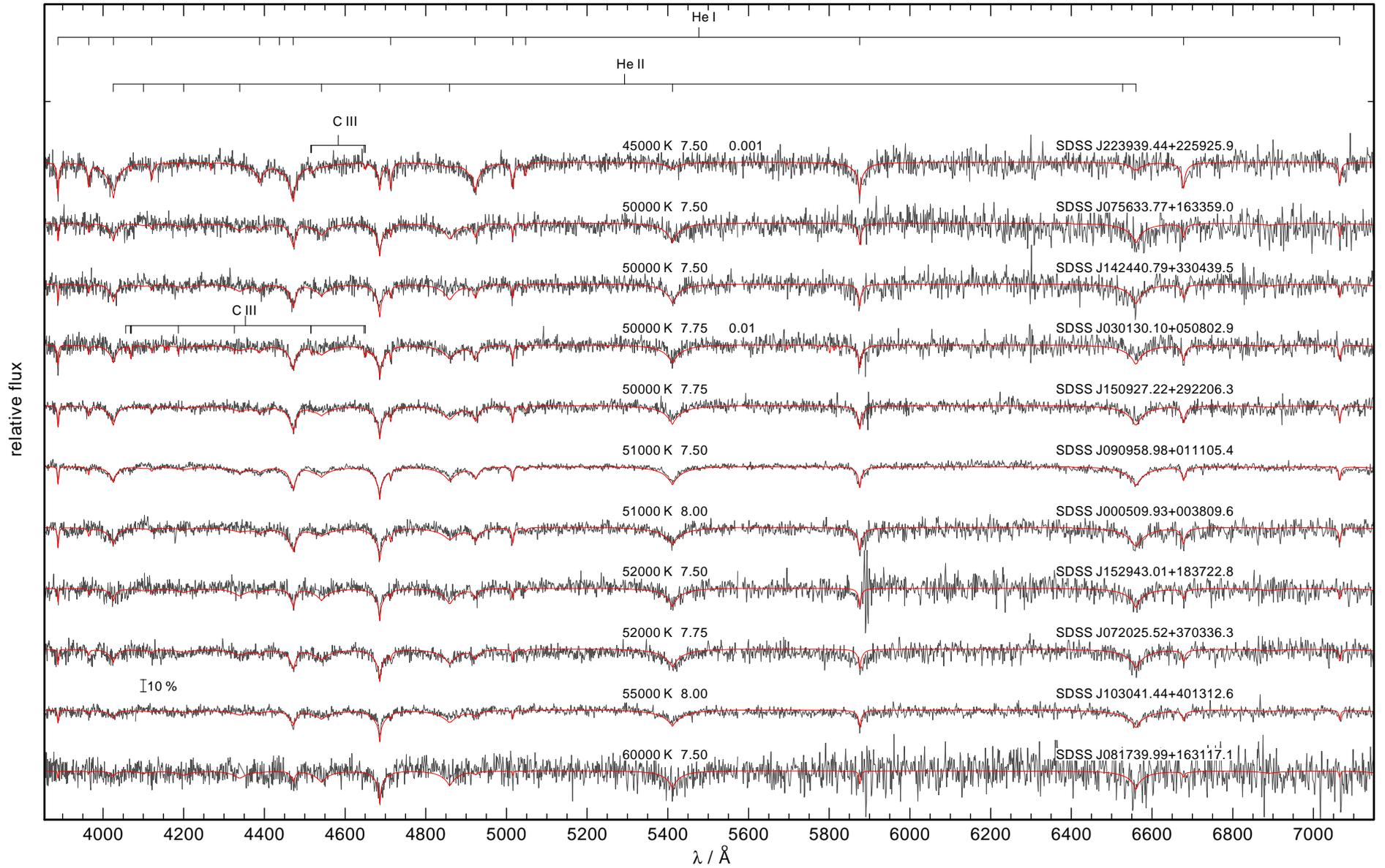}
  \caption{Spectra of new cool DO WDs together with best-fit models. Each spectrum is labeled with \Teff, \logg, the C abundance (mass fractions, where determined), and the SDSS name. The locations of photospheric lines are marked. The vertical bar indicates 10\,\% of the continuum flux.}
  \label{fig:do1}
\end{figure*}
\end{landscape}

\addtocounter{figure}{-1}

\begin{landscape}
\addtolength{\textwidth}{6.3cm} 
\addtolength{\evensidemargin}{-3cm}
\addtolength{\oddsidemargin}{-3cm}
\begin{figure*}
\includegraphics[trim=0 0 0 -120,height=20.0cm,angle=0]{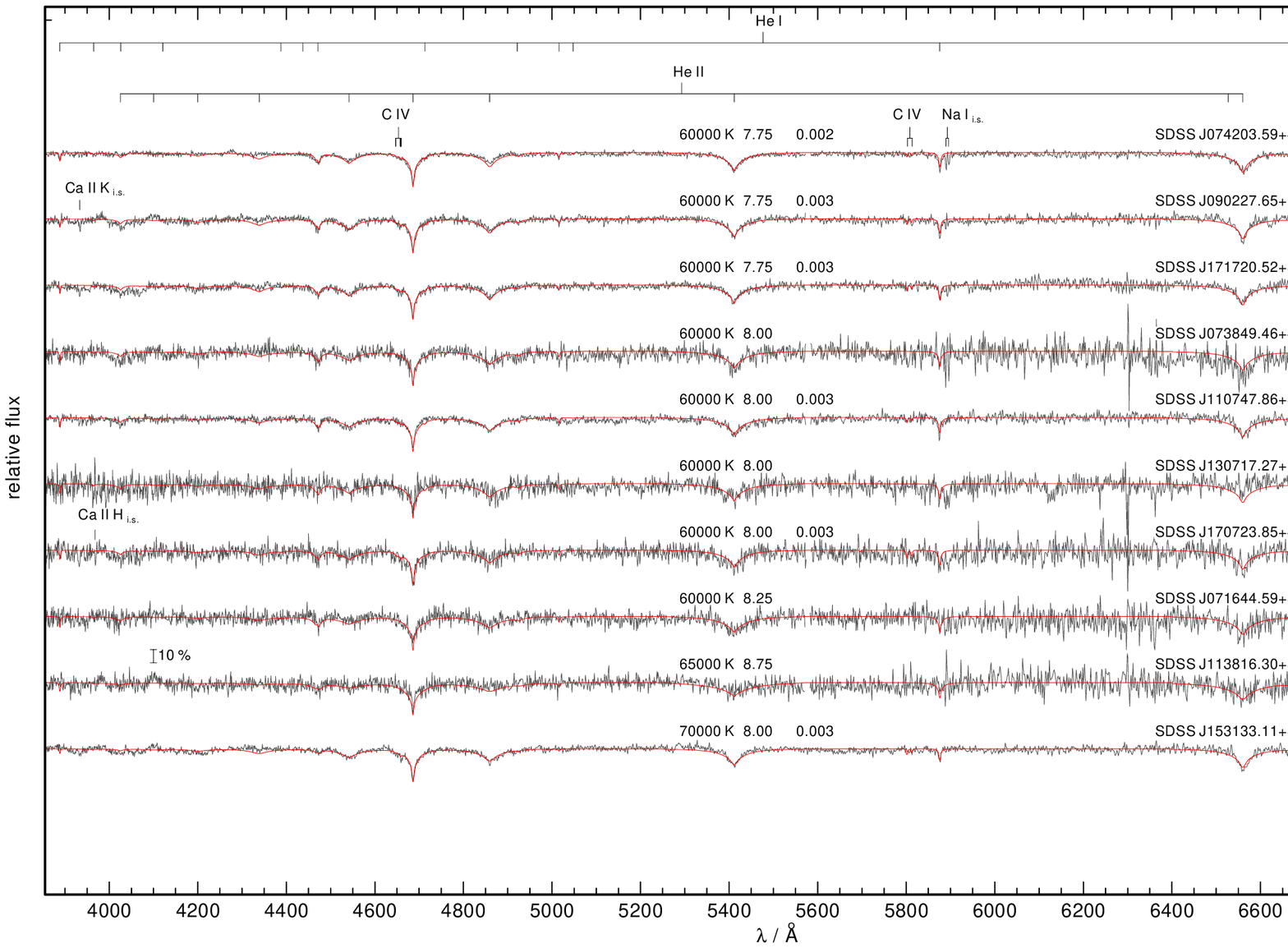}
  \caption{Continued.}
  \label{fig:do2}
\end{figure*}
\end{landscape}

\begin{landscape}
\addtolength{\textwidth}{6.3cm} 
\addtolength{\evensidemargin}{-3cm}
\addtolength{\oddsidemargin}{-3cm}
\begin{figure*}
\includegraphics[trim=0 0 0 -120,height=20.0cm,angle=0]{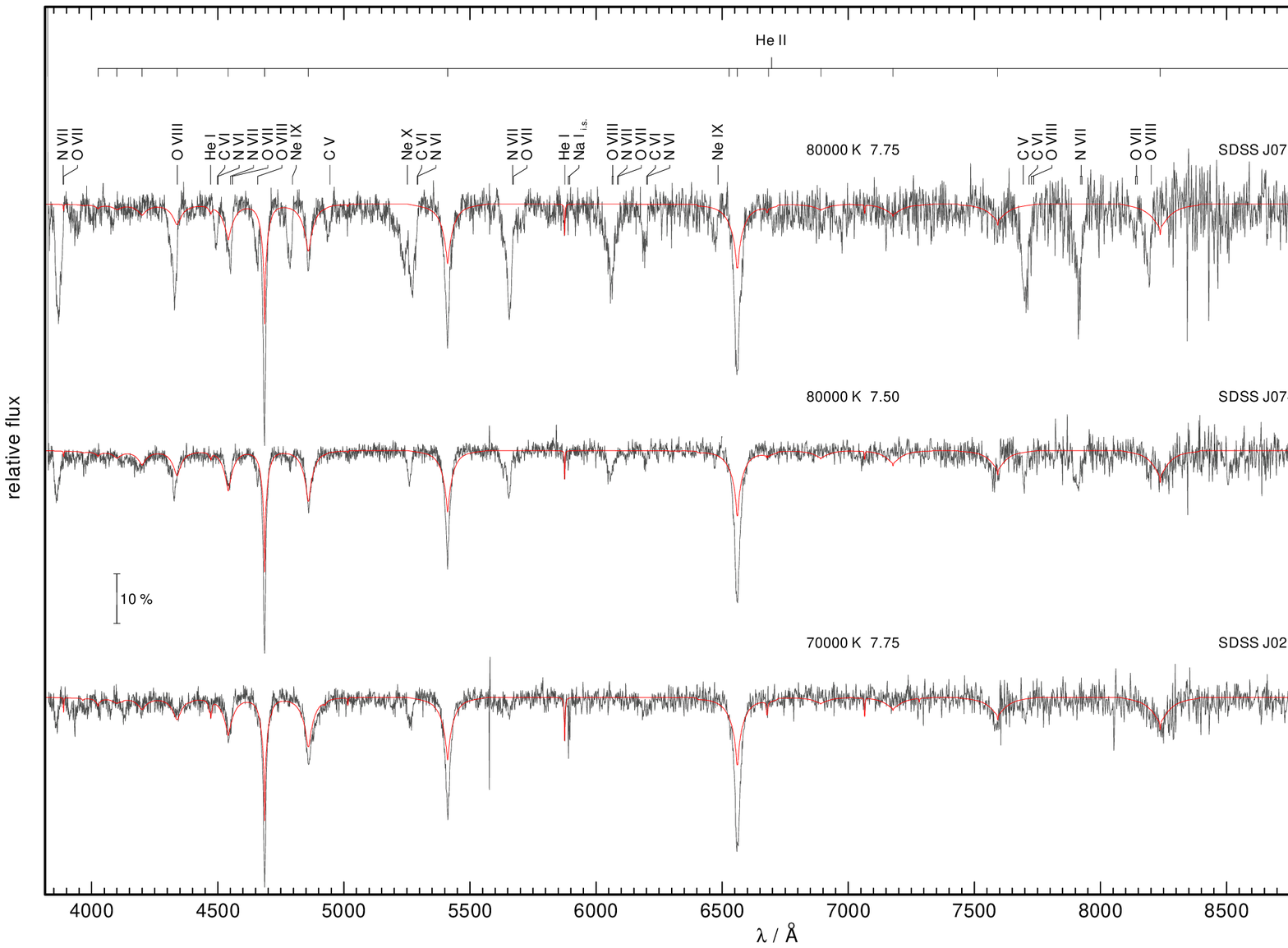}
  \caption{DO WDs with ultrahigh excitation absorption lines. Each spectrum is labeled with \Teff, \logg, and the SDSS name. The locations of photospheric lines and ultrahigh excitation features are marked. The vertical bar indicates 10\,\% of the continuum flux.}
  \label{fig:do3}
\end{figure*}
\end{landscape}

\begin{table*}[ht]
\caption{Parameters of the new cool DO WDs. C abundances are given as mass fraction. uhei indicates objects with ultrahigh excitation features.}
\label{tab:parameters}
\renewcommand{\tabcolsep}{1.5mm}
\begin{tabular}{l l l r@{.}l l l l l} 
SDSS name            & \Teff            & \logg           & \multicolumn{2}{l}{\,\,\,C}  & $M$             & $g^a$      & Remarks \\ 
                     &  [kK]            & [cm\,/\,s$^2$]           &  \multicolumn{2}{c}{}        & [\Msol]       & [mag] & \\ 
\hline 
\noalign{\smallskip}
\object{J000509.93+003809.6}\xspace  & 51$^{+2}_{-2}$    &  8.00$\pm$0.25   &  $<$\,0&003 &  0.64$^{+0.14}_{-0.10}$    & 18.181    &\\
\noalign{\smallskip}                                                                                       
\object{J020127.20+234952.7}\xspace  & 70$^{+10}_{-10}$  &  7.75$\pm$0.50    &  $<$\,0&002 &   0.58$^{+0.11}_{-0.06}$  & 16.633   & HS\,0158+2335, uhei\\
\noalign{\smallskip}                                                                                      
\object{J030130.10+050802.9}\xspace  & 50$^{+2}_{-1}$    &  7.75$\pm$0.25   & 0&01$^{+0.01}_{-0.005}$ &  0.54$^{+0.10}_{-0.07}$  & 18.392   &   \\
\noalign{\smallskip}                                                                                      
\object{J071644.59+395808.9}\xspace  & 60$^{+10}_{-5}$   &  8.25$\pm$0.50    & $<$\,0&05   &  0.75$^{+0.15}_{-0.09}$  & 18.591   &  \\
\noalign{\smallskip}                                                                                       
\object{J071702.72+395323.6}\xspace  & 80$^{+10}_{-10}$  &  7.75$\pm$0.50    &  $<$\,0&003  &  0.60$^{+0.09}_{-0.07}$  & 16.295   &  HS\,0713+3958, uhei\\
\noalign{\smallskip}                                                                                       
\object{J072025.52+370336.3}\xspace  & 52$^{+2}_{-2}$    &  7.75$\pm$0.25   &  $<$\,0&01  &  0.54$^{+0.10}_{-0.07}$  & 18.271    &   \\
\noalign{\smallskip}                                                                                       
\object{J073849.46+485126.6}\xspace  & 60$^{+10}_{-5}$   &  8.00$\pm$0.50    &  $<$\,0&01  &  0.66$^{+0.26}_{-0.16}$  & 18.695  & \\
\noalign{\smallskip}                                                                                      
\object{J074203.59+493333.8}\xspace  & 60$^{+10}_{-5}$   &  7.75$\pm$0.25   & 0&002$^{+0.001}_{-0.001}$ &  0.56$^{+0.10}_{-0.06}$  & 16.707  &  \\
\noalign{\smallskip}                                                                                       
\object{J074725.15+651301.1}\xspace  & 80$^{+10}_{-10}$  &  7.50$\pm$0.50    & $<$\,0&002   &  0.53$^{+0.16}_{-0.08}$  & 15.414   & HS\,0742+6520, uhei\\
\noalign{\smallskip}                                                                                       
\object{J075633.77+163359.0}\xspace  & 50$^{+2}_{-2}$    &  7.50$\pm$0.25   & $<$\,0&005  &  0.49$^{+0.05}_{-0.04}$  & 18.891   & \\
\noalign{\smallskip}                                                                                      
\object{J081739.99+163117.1}\xspace  & 60$^{+10}_{-9}$   &  7.50$\pm$0.50    & $<$\,0&05   &  0.50$^{+0.06}_{-0.04}$  & 19.184  &\\
\noalign{\smallskip}                                                                                      
\object{J090227.65+125206.0}\xspace  & 60$^{+5}_{-5}$    &  7.75$\pm$0.25   & 0&003$^{+0.002}_{-0.001}$ &  0.56$^{+0.10}_{-0.06}$  & 17.049 & \\
\noalign{\smallskip}                                                                                     
\object{J090958.98+011105.4}\xspace  & 51$^{+2}_{-2}$    &  7.50$\pm$0.25   & $<$\,0&003  &  0.49$^{+0.05}_{-0.04}$  & 16.660  & \\
\noalign{\smallskip}                                                                                     
\object{J103041.44+401312.6}\xspace  & 55$^{+5}_{-3}$    &  8.00$\pm$0.25   & $<$\,0&005  &  0.65$^{+0.13}_{-0.16}$  & 17.348   &  \\
\noalign{\smallskip}                                                                                      
\object{J110747.86+383550.8}\xspace  & 60$^{+5}_{-5}$    &  8.00$\pm$0.25   & 0&003$^{+0.002}_{-0.001}$  &  0.66$^{+0.12}_{-0.10}$  & 17.173    & \\
\noalign{\smallskip}                                                                                       
\object{J113816.30+382635.1}\xspace  & 65$^{+10}_{-10}$ &  8.75$\pm$0.50    &  $<$\,0&01  &  1.08$^{+0.18}_{-0.29}$  & 18.080    &   \\ 
\noalign{\smallskip}                                                                                       
\object{J130717.27+004151.6}\xspace  & 60$^{+10}_{-10}$  &  8.00$\pm$0.50    &  $<$\,0&01  &  0.66$^{+0.26}_{-0.16}$  & 17.188   & \\
\noalign{\smallskip}                                                                                      
\object{J142440.79+330439.5}\xspace  & 50$^{+2}_{-2}$    &  7.50$\pm$0.25   & $<$\,0&005  &  0.49$^{+0.05}_{-0.04}$  & 18.572  &\\
\noalign{\smallskip}                                                                                       
\object{J150927.22+292206.3}\xspace  & 50$^{+2}_{-1}$    &  7.75$\pm$0.25   & $<$\,0&003  &  0.54$^{+0.10}_{-0.07}$  & 17.321   & \\
\noalign{\smallskip}                                                                                       
\object{J152943.01+183722.8}\xspace  & 52$^{+3}_{-3}$    &  7.50$\pm$0.25   &  $<$\,0&005 &  0.49$^{+0.05}_{-0.04}$  & 18.432  &  \\
\noalign{\smallskip}                                                                                       
\object{J153133.11+343327.5}\xspace  & 70$^{+10}_{-5}$   &  8.00$\pm$0.25   & 0&003$^{+0.002}_{-0.001}$     &  0.69$^{+0.09}_{-0.11}$  & 15.834  &  \\ 
\noalign{\smallskip}                                                                                      
\object{J170723.85+450009.9}\xspace  & 60$^{+10}_{-5}$   &  8.00$\pm$0.50    & 0&003$^{+0.002}_{-0.001}$     &  0.66$^{+0.26}_{-0.16}$  & 18.644  &\\
\noalign{\smallskip}                                                                                      
\object{J171720.52+373605.9}\xspace  & 60$^{+5}_{-5}$    &  7.75$\pm$0.25   & 0&003$^{+0.002}_{-0.001}$     &  0.56$^{+0.10}_{-0.06}$  & 17.880  & \\
\noalign{\smallskip}                                                                                       
\object{J223939.44+225925.9}\xspace  & 45$^{+1}_{-2}$    &  7.50$\pm$0.25   & 0&001$^{+0.002}_{-0.0009}$     &  0.47$^{+0.05}_{-0.05}$  & 18.248 &  \\
\noalign{\smallskip}
\hline
\end{tabular}
\tablefoot{~\\
\tablefoottext{a}{$g$ magnitudes were taken from \cite{Ahnetal2012}.}
}
\end{table*}

\section{Spectral analysis}
\label{sect:analysis}

We used the T{\"u}bingen NLTE model-atmosphere package 
\citep[TMAP\footnote{\url{http://astro.uni-tuebingen.de/~TMAP}},][]{werneretal2003,rauchdeetjen2003} 
to compute non-LTE, plane-parallel, fully metal-line-blanketed model atmospheres 
in radiative and hydrostatic equilibrium. The model atoms for this analysis were taken
from the T{\"u}bingen model-atom database TMAD\footnote{\url{http://astro.uni-tuebingen.de/~TMAD}}. 
To calculate synthetic line profiles, we used Stark line-broadening tables provided by \cite{Barnard1969} for \Ionww{He}{1}{4026, 4388, 4471, 4921}, 
\cite{Barnard1974} for \Ionww{He}{1}{4471} and \cite{Griem1974} for all other \ion{He}{I} lines, and for \ion{He}{II} and \ion{C}{IV} we used the tables 
provided by \cite{Schoening1989} and \cite{Schoening1993}.  To account for the spectral resolution of the observations, synthetic spectra 
were convolved with Gaussians ($FWHM = 2.5$\AA). All observed spectra were shifted to rest wavelengths by applying radial-velocity 
corrections by centroiding the \Ion{He}{1} and \Ion{He}{2} lines.\\
First, we calculated a pure-He model grid, spanning from \Teffw{45-90} (in steps of 5\,kK) and \loggw{6.75-8.75} 
(in steps of 0.25). Because the \ion{He}{I}\,/\,\ion{He}{II} ionization equilibrium is very sensitive 
around \Teffw{50}, we refined the model grid from \Teffw{43-55} to 1\,kK steps. We reproduced all He lines and chose the best-fit models by visual 
comparison with the whole rectified observed WD spectra. To determine \Teff, we used the \ion{He}{I}\,/\,\ion{He}{II} ionization equilibrium, and for the 
\logg\ determination the wings of the He lines.\\
Then, we also included C ($C=0.0001, 0.0005$, $C=0.001-0.01$, in $0.001$ steps, and $C=0.01-0.05$, in $0.01$ steps, by mass) into our best-fit models 
to derive C abundances by fitting all identified C lines (\se{sect:observation}). Upper limits were derived by test models where the respective lines in 
the model contradicted the non-detection of the lines in the observation (at the abundance limit).
For five stars (J0902, J1107, J1531, J1707, and J1717), we derived $C=0.003$, which is slightly supersolar 
(solar abundance according to \citealt{asplundetal2009}). For J0742 we found $C=0.002$ (about solar). Two of our objects are
the coolest DO WDs ever discovered that still show a considerable amount of C. J0301 (\Teffw{50}) was found to have the highest 
C abundance in our sample ($C=0.01$, about four times solar), and for J2239 (\Teffw{45}) we found $C=0.001$ (slightly subsolar).\\
The three DO WDs with uhei features clearly show \Ionw{He}{1}{5875.7}; the high S/N of the spectra of J0747 and J0201 also
allowed the identification of \Ionw{He}{1}{4471.5}. We used these lines to constrain \Teff\ and found \Teffw{70} for J0201 and \Teffw{80} 
for J0717 and J0747. For the surface gravity of these objects we estimate \loggw{7.75} (for J0717 and J0201) and 
\loggw{7.50} (for J0747). For lower values of \logg, the Pickering lines below 4500\,\AA\ become too strong. Upper limits 
for \logg\ were derived using \Ionww{He}{2}{4542, 4860, 8237}, which form deep in the atmosphere and might thus be least affected 
by the process that creates the too-deep \ion{He}{II} lines. We summarize the results of our analysis in Table~\ref{tab:parameters}.\\
The additional opacities of C do not affect the theoretical \ion{He}{I} and \ion{He}{II} line profiles. 
We also investigated a possible impact of line blanketing by iron-group elements. For that, we calculated two test models with 
\Teffw{60, 80}, and \loggw{7.0}. Iron-group elements were included with a generic model atom \citep{rauchdeetjen2003} containing the elements 
Ca, Sc, Ti, V, Cr, Mn, Fe, Co, and Ni at solar abundance values. For the \Teffw{60} model the ionization stages {\sc iv} $-$ {\sc vii} 
were considered and for the \Teffw{80} model the ionization stages {\sc v} $-$ {\sc ix}. The model atom was calculated via the 
T{\"u}bingen iron-group opacity interface TIRO\footnote{\url{http://astro.uni-tuebingen.de/~TIRO}} \citep{ringatPhD2013}. This has recently been 
developed in the framework of the Virtual Observatory (\emph{VO}\footnote{\url{http://www.ivoa.net}}) and is provided as a registered 
service by the \emph{German Astrophysical Virtual Observatory}\footnote{\url{http://www.g-vo.org}}. We found that for the pure-He models, 
the central depression of \Ionw{He}{2}{4686} is stronger than in the model that includes the iron-goup elements (42\% stronger in the \Teffw{60} 
model and 10\% stronger in the \Teffw{80} model). However, the line profiles of all the other \ion{He}{I} and \ion{He}{II} lines remain unaffected.

\section{Discussion}
\label{sect:discussion}

\subsection{Hot-wind DO WDs}
\label{subsect:uhei}

The discovery of uhei lines and unusually deep \Ion{He}{2} lines in the spectrum of J0747 makes it the eleventh member 
of the so-called hot-wind DO WDs. This class so far included eight other DO WDs: \object{SDSS\,J105956.00+404332.4}\xspace \citep{werneretal2014}, 
\object{SDSS\,J151026.48+610656.9}, \object{SDSS\,J025403.75+005854.4}\xspace \citep{huegelmeyeretal2006}, \object{HS\,0158+2335}\xspace, \object{HS\,0727+6003}\xspace, 
\object{HS\,2027+0651}\xspace \citep{Dreizleretal1995}, \object{HE\,0504-2408}\xspace, \object{HS\,0713+3958}\xspace \citep{werneretal1995}, one 
PG\,1159 star: \object{SDSS\,J121523.09+120300.8}\xspace \citep{huegelmeyeretal2006}, and one DAO WD \object{HS\,2115+1148}\xspace \citep{Dreizleretal1995}. 
In addition, there are five other DO WDs known that also show too deep \Ion{He}{2} lines that they cannot be fitted by any model, but have no clear uhei features in their 
spectra: \object{SDSS\,J082134.95+173919.40}\xspace, \object{SDSS\,J082724.44+585851.68}\xspace, \object{SDSS\,J094722.49+101523.62}\xspace, 
\object{SDSS\,J151215.72+065156.34}\xspace \citep{werneretal2014}, and \object{HE\,1314+0018}\xspace \citep{Werneretal2004, huegelmeyeretal2006}. 
\cite{werneretal2014} speculated that the same unknown process is at work here, affecting the \Ion{He}{2} lines, but failing to 
generate the strong uhei lines.\\
An attempt was made by \cite{werneretal1995} to explain the uhei absorptions lines by extremely hot, static, plane-parallel 
non-LTE model atmospheres. The results showed that the observed spectra cannot have a photospheric origin assuming such an 
extreme \Teff. Since the strongest of the uhei lines often show a very broad blue wing, it is believed that they are 
formed in a hot stellar wind that is optically thick in these transitions along the line of sight toward the stellar disk.\\
From constraining atmospheric parameters for these stars with the method described in \se{sect:analysis}, 
we cannot find any correlation of the temperature (e.g., the presence of \Ion{He}{1} lines), and 
the presence or strength of uhei features in these DO WDs.\\
The fraction of DO WDs showing this phenomenon is significant (19\% show too deep \Ion{He}{2} lines, 11\% show additional uhei features) 
and awaits an explanation.\\

\subsection{C abundances}
\label{subsect:C}

\begin{figure*}[ht]
  \resizebox{\hsize}{!}{\includegraphics{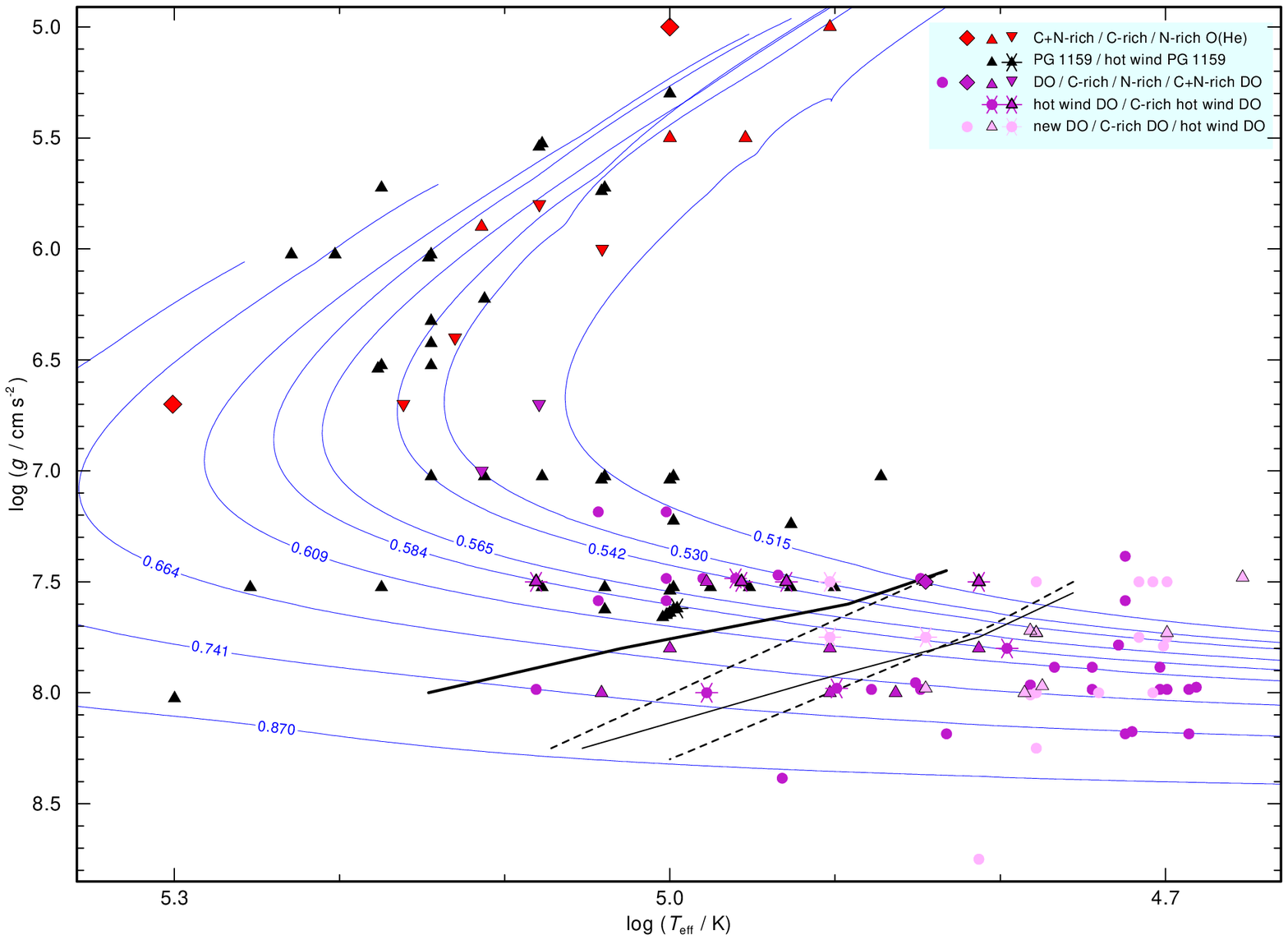}}
  \caption{Locations of O(He) stars (red, \citealt{Reindletal2014, werneretal2014, wassermannetal2010, DeMarco2014}), PG\,1159 stars (black, Geier et al. in prep.; \citealt{werneretal2014, Gianninas2010, wernerherwig2006, Schuh2008}) and DO WDs (purple, this work, \citealt{werneretal2014, mahsereci2011, huegelmeyeretal2006, dreizlerwerner1996, Nageletal2006}) in the log \Teff\ -- \logg\ plane compared to  VLTP post-AGB (solid lines) evolutionary tracks (labeled with stellar masses in $M_\odot$) of \cite{althausetal2009}. The black solid and dashed lines indicate theoretical wind limits for abundance changes as predicted by \cite{UnglaubBues2000} and discussed in \se{subsect:C}.}
  \label{fig:vltp}
\end{figure*}

\begin{figure}[ht]
  \resizebox{\hsize}{!}{\includegraphics{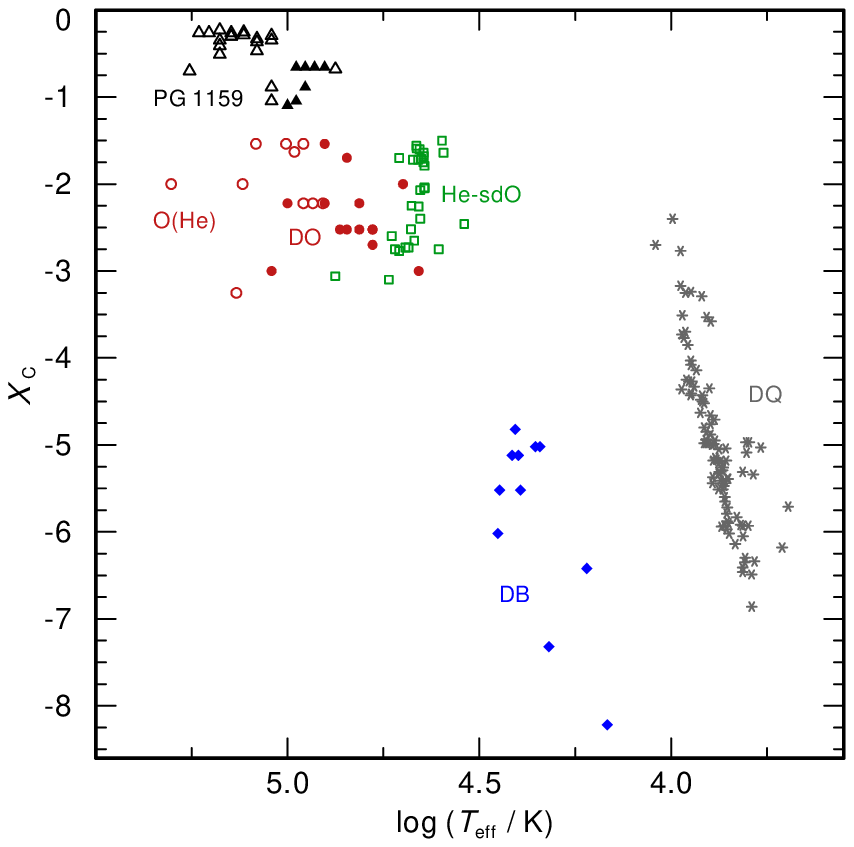}}
  \caption{Carbon abundances (in logarithmic mass fractions) before and along the non-DA WD cooling track. Luminous PG\,1159 stars with \Teff$> 120$\,kK and \logg$< 7.5$ are represented by open, black triangles, PG\,1159 stars close to the wind limit with \Teff$\leq 120$\,kK and \logg$\geq 7.5$ by filled, black triangles, O(He) stars and DO WDs before the wind limit by open, red circles, DO WDs beyond the wind limit by filled, red circles, DB WDs by blue rhombs, and DQ WDs by gray stars.}
  \label{fig:CTeff}
\end{figure}

The detection of C in cool DO WDs is of particular interest in studying the chemical evolution of WDs and to place constraints on 
possible progenitors and successors of DO WDs.\\
The chemical evolution of hot WDs in the presence of diffusion and mass loss was studied by \cite{UnglaubBues2000}. Using the luminosity 
dependence of the mass-loss rate of \cite{bloecker1995}, they predicted that until \Teffw{65} the C abundance of a 
0.529\,\Msol\ star is reduced only by a factor of two. For stars with higher mass this occurs somewhat earlier, as indicated by the 
upper dashed line in Fig.\,\ref{fig:vltp}. This figure also shows the locations of all analyzed DO WDs, PG\,1159, and O(He) stars compared with 
VLTP evolutionary tracks from \cite{althausetal2009}. As a consequence of increasing gravity and decreasing mass-loss rates, the effect of 
gravitational settling becomes more and more apparent. When the star has reached the lower dashed line in Fig.\,\ref{fig:vltp}, the C 
abundance is expected to be reduced by a factor of ten. If the dependence of mass loss on the chemical composition is considered, an even sharper 
transition of PG\,1159 stars into DO WDs is expected. We also show in Fig.\,\ref{fig:vltp} the corresponding theoretical wind 
limit for PG\,1159 stars in which gravitational settling overcomes radiation-driven mass loss (thin solid line). 
The thick solid line in Fig.\,\ref{fig:vltp} corresponds to a ten times lower mass-loss rate and relates to the \textit{observed} 
PG\,1159 wind limit because no PG\,1159 star is observed beyond that region. Assuming a lower mass-loss rate, however, also implies that the 
decrease of C abundances in DO WDs occurs earlier. \cite{UnglaubBues2000} claimed that the co-existence of PG\,1159 stars and DO WDs with 
various compositions and similar stellar parameters does not contradict an evolutionary link. Depending on their mass-loss rates, some objects 
may evolve into DO WDs somewhat earlier, others later.\\
Figure\,\ref{fig:CTeff} illustrates the observed (logarithmic) C abundances as a function of \Teff for PG\,1159 stars (values taken from Geier et al. in prep.; 
\citealt{wernerrauch2014, werneretal2014, Schuh2008, wernerherwig2006, huegelmeyeretal2006}), O(He) stars \citep{Reindletal2014, 
werneretal2014, wassermannetal2010}, C-rich He-sdOs \citep{Nemeth2012, hirsch2009}, DO WDs, (this work, \citealt{werneretal2014,  
huegelmeyeretal2006, dreizlerwerner1996}), DB WDs \citep{Provencal1996, Provencal2000, Dufour2002, Petitclerc2005, Desharnais2008, Koester2014} and 
DQ WDs \citep{Koester2006, Dufour2005}. Although there are some He-rich luminous PG\,1159 stars, the vast majority of the luminous,
 that is, less evolved, 
PG\,1159 stars (black, open triangles) display $C \approx 0.5$. Obviously, PG\,1159 stars close to the wind limit (\Teff\,$\leq 120$\kK\,\,and 
\logg$\geq 7.5$, black, filled triangles) display the lowest C abundances ($C \leq 0.22$, by mass), which means that they
are lower by about a factor of two than
the average of the luminous stars. This observational fact indeed supports the theory of advancing gravitational settling.\\
On the other hand, the typical C abundances in DO WDs (0.001-0.01 by mass) do not change along the WD cooling track and are much closer to 
the C abundances observed in C-rich O(He) stars (typically 0.1--0.03 by mass, \citealt{werneretal2014, Reindletal2014, wassermannetal2010}) than to 
those of PG\,1159 stars. The discovery of a significant amount of C in the atmosphere of J0301, and still some detectable C in the 
atmosphere of J2239, emphasizes that gravitational settling might work less well than predicted by theory in the DO WD cooling region. 
According to \cite{UnglaubBues2000}, the C abundances of J0301 and J2239 should already have dropped far below the detectable limit. \cite{Dreizler1999}
also found no evidence for gravitational settling. By analyzing HST spectra of DO WDs ranging from 
\Teffw{50-100}, he found that DO WDs can best be fitted with chemically homogeneous models, whereas the stratified models significantly 
deviated from the observations.\\
It is also interesting to note that the hot-wind DO WDs are located around the PG\,1159--DO transition region predicted by \cite{UnglaubBues2000}. 
Given that the cause for their spectral anomaly is a fast wind, this would additionally stress that some mass-loss must still
be occurring in DO 
WDs.\\
The observed C abundances found in DO WDs are also very similar to those observed in C-rich He-sdO stars. A non-LTE 
analysis of a large sample of these stars revealed C abundances in the range of 0.001-0.03 \citep{Nemeth2012, hirsch2009}.
Especially J2239, for which we derived a mass of 0.47\,\Msol, might be a good candidate for a successor of a C-rich He-sdO star.\\
Traces of C (0.9-9.5 $ \times 10^{-6}$, by mass) are also found in some hot DB WDs 
\citep{Provencal1996, Provencal2000, Dufour2002, Petitclerc2005, Desharnais2008, Koester2014} 
and cannot be easily explained either by any physical processes currently thought to operate in the envelopes of DB stars. Numerical simulations of 
\cite{Fontaine2005} showed that assuming a weak stellar wind of about $10^{-13}$\Msol/yr would sufficiently slow down the settling of C. 
However, their wind model has no physical basis and is not compatible with the thin radiatively driven winds discussed by \cite{UnglaubBues2000} at much 
higher \Teff. Moreover, the mass-loss rates are far below the detection limit.\\
The monotonic decrease of the C abundance with decreasing temperature in DQ stars, uncovered by \cite{Dufour2005} and confirmed by \cite{Koester2006}, 
can be explained in terms of the dredge-up model developed by \cite{Pelletier1986}. They predicted that the highest contamination occurs at around \Teffw{12}, 
approximately the \Teff at which the base of the He convection zone reaches its highest depth. Below this \Teff, C pollution decreases with further cooling, 
mainly because C sinks back into the star as a result of its partial recombination. The tight observational sequence found by \cite{Dufour2005} and 
\cite{Koester2006} allowed \cite{Brassard2007} to pin down the masses of the He-dominated envelopes in DQ stars ($10^{-2}$ to $10^{-3.75}$\,\Msol), which agrees with (V)LTP models. Hence, they reaffirmed the natural connection between PG\,1159 stars, DO, DB, and DQ WDs. A completely different result was 
obtained by \cite{Scoccola2006}. They found that PG 1159 stars cannot be related to any DQ WDs with low C abundances and instead suggested that the latter 
could be successors of RCB stars or C-poor post-extreme horizontal branch (EHB) stars. In that sense, we propose that  O(He) stars should also be investigated as 
possible progenitors of DQ WDs. It is important to note,
however, that \cite{Scoccola2006} failed to reproduce the decrease of the C abundance with 
decreasing temperature in DQ stars and instead found an \textit{increase} of the C abundance with decreasing temperature.

\begin{table}[ht]\centering
\caption{Identification of ultrahigh excitation features in J0717, J0747, and J0201. 
$\times$ denotes that these lines were already identified in J0717 by \cite{werneretal1995} and in 
J0201 by \cite{Dreizleretal1995}, while $\star$ denotes newly identified lines.}
\label{tab:uhei}
\renewcommand{\tabcolsep}{1.5mm}
\begin{tabular}{l l c c c} 
\hline 
\noalign{\smallskip}
ion           & $\lambda$\,/\,\AA & J0717 & J0747 & J0201 \\
\hline 
\noalign{\smallskip}
\Ion{C}{5}    & 4945$^4$          &   $\times$      &         &       \\ 
              & 7694$^2$          &   $\star$       & $\star$ &  $\star$  \\ 
\Ion{C}{6}    & 4499$^5$          &   $\times$      &         &       \\ 
              & 5291$^5$          &   $\times$      & $\star$ &       \\ 
              & 6201$^5$          &   $\times$      & $\star$ &       \\
              & 7717$^2$          &   $\star$       & $\star$ &       \\
\Ion{N}{6}    & 4501$^1$          &   $\times$      &         &       \\ 
              & 5293$^1$          &   $\times$      & $\star$ &  $\times$    \\ 
              & 6204$^1$          &   $\times$      & $\star$ &  $\times$    \\ 
\Ion{N}{7}    & 3887$^5$          &   $\times$      & $\star$ &  $\times$    \\ 
              & 4550$^5$          &   $\times$      & $\star$ &  $\times$    \\ 
              & 5669$^5$          &   $\times$      & $\star$ &  $\times$    \\ 
              & 6085$^5$          &   $\times$      & $\star$ &  $\times$    \\ 
              & 7921$^1$          &   $\star$       & $\star$ &       \\ 
              & 7926$^1$          &   $\star$       & $\star$ &       \\ 
\Ion{O}{7}    & 3889$^1$          &   $\times$      & $\star$ &  $\times$    \\ 
              & 4558$^1$          &   $\times$      & $\star$ &  $\times$    \\ 
              & 5673$^1$          &   $\times$      & $\star$ &  $\times$    \\ 
              & 6088$^1$          &   $\times$      & $\star$ &  $\times$    \\ 
              & 8139$^3$          &   $\star$       &         &       \\ 
              & 8146$^3$          &   $\star$       &         &       \\ 
\Ion{O}{8}    & 4340$^5$          &   $\times$      & $\star$ &       \\ 
              & 4658$^5$          &   $\times$      & $\star$ &  $\times$    \\ 
              & 6064$^5$          &   $\times$      & $\star$ &  $\times$    \\ 
              & 6068$^5$          &   $\times$      & $\star$ &  $\times$    \\ 
              & 7726$^1$          &   $\star$       & $\star$ &       \\ 
              & 7736$^1$          &   $\star$       & $\star$ &  $\times$    \\ 
              & 8202$^1$          &   $\star$       & $\star$ &       \\ 
\Ion{Ne}{9}   & 4797$^1$          &   $\times$      & $\star$ &  $\star$  \\ 
              & 6484$^1$          &   $\times$      & $\star$ &  $\times$    \\ 
\Ion{Ne}{10}  & 5252$^1$          &   $\times$      & $\star$ &  $\times$    \\ 
\noalign{\smallskip}
\hline
\end{tabular}
\tablefoot{~\\
\tablefoottext{1}{hydrogenic value}
\tablefoottext{2}{\cite{Engstrom1992}}
\tablefoottext{3}{\cite{Accad1971}}
\tablefoottext{4}{\cite{Moore1970}}
\tablefoottext{5}{Garcia \& Mack (1965)}
}
\end{table}

\subsection{Mass distribution}
\label{subsect:mass}

\begin{figure}[ht]
  \resizebox{\hsize}{!}{\includegraphics{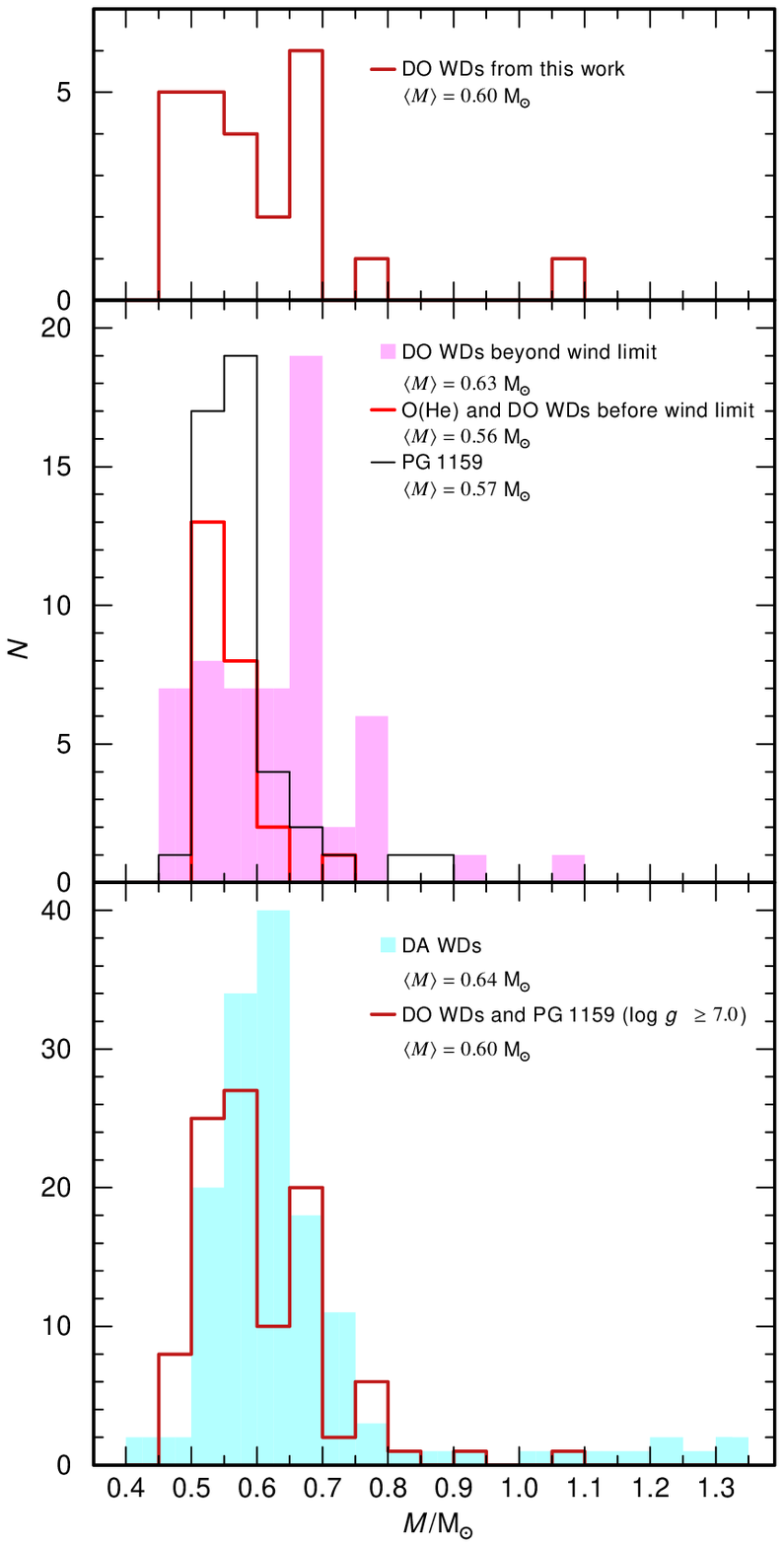}}
  \caption{Upper panel: Mass distribution of all DO WDs from our sample. Middle panel: Mass distribution of PG\,1159 stars (thin black line), O(He) stars and DO WDs before the wind limit (thick red line), and DO WDs beyond the wind limit (gray area, light pink in the online version). Lower panel: Mass distribution of the hot (\Teff\,$\geq$ 45\,kK) DA WDs  (gray area, light blue in the online version) from the sample of \cite{Gianninas2011} compared with the mass distribution of non-DA WDs (PG\,1159 and DO WDs with \logg\,$\geq$ 7, thick red line).}
  \label{fig:MassDist}
\end{figure}

We derived the masses of the new cool DO WDs by comparing their positions in the log \Teff\ -- \logg\ plane with VLTP evolutionary tracks from 
\citet[][Fig.\,\ref{fig:vltp}]{althausetal2009}. With \Teffw{65} and \loggw{8.75}, J1138 clearly lies outside of the region covered by the VLTP 
tracks. Considering J1138 as CO-WD, we used CO-WD evolutionary tracks of \cite{Wood1994}, which extend up to 1.10\Msol. We found $M=1.09$\,\Msol\ 
with a lower limit of 0.79\Msol. Considering J1138 as an ONe-WD we used ONe-WD tracks of \cite{Althaus2005b} and derive $M=1.07$\,\Msol\ with an 
upper limit of 1.26\,\Msol.
\cite{Lauetal2012} showed that because of instabilities in the late thermally pulsing-AGB phase of massive AGB stars 
with zero-age main-sequence masses of 7-10\,\Msol\, and envelope masses of about 1-2\,\Msol,\ most of the envelope of these stars can be ejected. 
The outcome might be a central star (CS) of around 1\,\Msol\ with a relatively massive planetary nebula (PN), like the so-far unique He-rich 
CS N66 (\object{SMP\,83}), whose luminosity corresponds to a core mass of about 1.2\,\Msol\ \citep{Hamann2003}. J1138 could be a successor 
of such a star. The very low frequency of massive DO WDs could be consistent with the fact that not all AGB stars going 
through such an ejection mechanism would eject all of their envelope and thus would instead become a massive DA WD.\\
The derived masses together with their errors (as resulting from the uncertainties in \Teff\ and \logg) are given in Table~\ref{tab:parameters}. 
For J1138, we adopted the mean value given by both sets of tracks.\\
In the upper panel of Fig.\,\ref{fig:MassDist}, we show the mass distribution of the DO WDs of our sample. We derived a mean mass of 
$\langle M \rangle = 0.60$\,\Msol\ with a standard deviation of $\sigma = 0.13$\,\Msol. In the middle panel of Fig.\,\ref{fig:MassDist} we show the 
mass distribution of all analyzed PG\,1159 stars, O(He) stars, and DO WDs before the observed wind limit\footnote{Since no gravitational settling is predicted for DO 
WDs before the wind limit, we placed them in the O(He) group.} and DO WDs beyond the wind limit. To avoid systematic errors introduced by using different evolutionary 
tracks, we derived the masses for PG\,1159 and O(He) stars by additionally using the VLTP tracks from \cite{althausetal2009}. 
The mean mass of the O(He) stars and DO WDs before the wind limit ($\langle M \rangle =0.56$\,\Msol, with $\sigma = 0.04$\,\Msol) is slightly lower 
than the mean mass of the PG\,1159 stars ($\langle M \rangle=0.57$\,\Msol, $\sigma = 0.08$\,\Msol). The mean mass of DO WDs beyond the wind limit is 
slightly higher ($\langle M \rangle =0.63$\,\Msol, $\sigma = 0.11$\,\Msol) than those of the objects before the wind limit. This trend continues in the mass 
distribution of DB WDs. Using only the spectra with S/N $\geq 15$ out of their sample of 923 DB stars, \cite{Kleinman2013} found a mean mass of 
$\langle M \rangle =0.685\pm0.013$\,\Msol. By restricting this sample to just those hotter than \Teffw{16}, they found a slightly lower mean mass of 
$\langle M \rangle =0.676\pm0.014$\,\Msol. \cite{Bergeron2011} presented a detailed analysis of 108 DB WDs based on model atmosphere fits to high S/N 
optical spectroscopy. They derived an almost identical mean mass of 0.67\,\Msol\ for their sample.\\
The phenomenon of increasing mean mass with decreasing \Teff\ is not restricted to non-DA WDs. \cite{Gianninas2011} presented the spectroscopic analysis of over 
1100  bright ($V \geq 17.5$) DA white dwarfs. They found a mean mass of $\langle M \rangle = 0.661$\,\Msol, with $\sigma = 0.160$\,\Msol, but by dividing their sample 
into objects with \Teff$>$\,13\,kK and \Teff$<$\,13\,kK, they also found that the mean mass of the hot objects 
($\langle M \rangle = 0.638$\,\Msol, $\sigma = 0.145$\,\Msol) is significantly lower than the mean mass of the cool DA WDs 
($\langle M \rangle = 0.736$\,\Msol, $\sigma = 0.183$\,\Msol). \cite{Tremblayetal2013} showed that this problem can be solved using 3D hydrodynamical models 
to compute spectra for the cool DA WDs. 
In the lower panel of Fig.\,\ref{fig:MassDist} we therefore compare only the mass distributions of DA WDs with \Teff$>$\,45\,kK with those of the 
DO WDs and PG\,1159 stars with \logg$\geq 7.0$. We derived mean masses of both classes to $\langle M \rangle=0.64$\,\Msol\ 
(hot DA WDs) and $\langle M \rangle=0.60$\,\Msol\ (non-DA WDs). The slightly higher mean mass for DA WDs results from the larger number of massive WDs, while 
a larger number of non-DA WDs with $M<0.5$\,\Msol\ is known. Seven out of the 55 (about 13\%) DO WDs beyond the wind limit have $M<0.5$\,\Msol, thus 
they are most likely  post-EHB stars. The progenitors of these DO WDs might be He-sdO stars or low mass PG\,1159 stars, such
as \object{HS\,0704+6153}\xspace \citep{Dreizler1998}. 
However, we emphasize that the errors on the masses are still large and that only a better \logg determination can provide better constraints.\\
Another interesting feature that can be seen in the middle panel of Fig.\,\ref{fig:MassDist} is that the mass distribution of the DO WDs beyond the wind 
limit compared with the mass distribution of the objects before the wind limit strongly disagree. While the flat plateau around 0.6\,\Msol\ 
in the mass distribution of the DO WDs beyond the wind limit agrees with the mass distribution of PG\,1159 stars, O(He) stars, and 
DO WDs before the wind limit, two additional peaks at 0.675\,\Msol\ and 0.775\,\Msol\ can be seen. The significant difference in the mass distributions 
is confirmed by a Kolmogorov-Smirnov test. The result of this test is that the probability that both samples are taken from the same mass distribution is rather 
small ($<10^{-4}$). The shape of the mass distributions before and beyond the wind limit are also preserved if we divide the 
stars into objects younger and older than 0.4\,Myr, which approximately corresponds to a vertical line at \Teffw{80} \citep{althausetal2009}. This shows 
that the discrepancy in the mass distributions does not arise
because the wind limit is reached earlier for more massive stars. 
\cite{althausetal2009} noted that the mass distribution of young DO WDs differs considerably from that of the older WDs. \cite{werneretal2014} 
attributed the higher mean mass for DOs found by \cite{huegelmeyeretal2006} to a calibration problem in the SDSS DR4. However, we found that using only the 
DO WDs from our analysis, we also see a peak around 0.675\Msol. Thus we believe that this peak is real and not an artifact of a poor flux calibration.\\ 
This poses the question whether these two high-mass peaks display different input channels of DO WDs. \cite{althausetal2009} suggested that some DO WDs 
might result from evolutionary channels that do not involve PG\,1159 stars. For instance, they could be the result of post-merger evolution involving 
RCB, EHe, luminous He-sdO stars, and O(He) stars. \cite{Reindletal2014} found that using double He-WD merger tracks from \cite{zhangetal2012a} instead of 
VLTP tracks, the masses of O(He) stars are 0.07$-$0.16\,\Msol\ higher. This would indeed shift the mean mass of these stars closer to the higher mass 
peaks of DO WDs beyond the wind limit. Comparing the evolutionary tracks of \cite{zhangetal2012a} and \cite{althausetal2009},
we found that they differ significantly only in the luminous part (\Teff$>$\,100\,kK) in the log \Teff\ -- \logg\ plane. We again derived the masses 
of all O(He) stars and DO WDs before the wind limit using the tracks of \cite{zhangetal2012a} and found a mean mass of $\langle M \rangle=0.62$\,\Msol\ 
instead of the $\langle M \rangle=0.56$\,\Msol \ found using the tracks of \cite{althausetal2009}. 
Although a double He-WD merger can be excluded for some O(He) stars, we claim that a contribution of post-double He-WD mergers to the observed higher mass 
peaks in the mass distribution of the DO WDs beyond the wind limit is possible. We also stress that determining the mass through evolutionary tracks 
requires knowing the evolutionary history of an object, which is very unclear at least for the He-dominated objects.\\
Another possible origin of DO WDs with higher masses are H-deficient [WC] or [WN]-type central stars. A spectroscopic mass determination for these objects 
is, however, extremely difficult since \logg\ cannot be derived from photospheric lines. Wind emission lines do not depend on the first approximation of 
L and M. For the spectral analysis of these stars, it is common practice to assume standard values of \logg\  or $M$ (e.g., \loggw{6.0} or $M=0.6$\,\Msol), 
which are equivalent input parameters for wind codes. The two currently known [WN]-type central stars \ic \citep{miszalskietal2012} and \abe 
\citep{Todtetal2013, Frewetal2014} are in an evolutionary state similar to or even later than the O(He) stars, but show much stronger stellar winds. 
\cite{Reindletal2014} speculated that [WN] stars are O(He) stars, but with higher masses and hence higher luminosity, which could explain the higher
 mass-loss rates of [WN] stars.

\subsection{Ratio of DAs to non-DAs  for hot WDs}
\label{subsect:ratio}

The sample of \cite{Gianninas2011} comprised 131 DA WDs with \Teff$>$\,45\,kK, while there are 49 PG\,1159 stars with \logg$\geq 7.0$, and DO WDs 
(hot non-DA WDs) known with $V \leq 17.5$. This would lead to a ratio of hot DAs to non-DAs of 2.7. However, neither sample can be considered complete. 
To correctly calculate the ratio of hot DAs to non-DAs, it is necessary to compare samples with the same the magnitude limit, sky 
coverage, and completeness. No systematic search for hot DA WDs in the SDSS DR 10 was presented so far, therefore we restricted ourselves to the SDSS DR7 
spectroscopic sample. Since we compare objects in the same \Teff\ range, incompleteness should affect both subclasses in the same manner. Atmospheric parameters 
derived by \cite{Kleinman2013} are only based on LTE models, and thus they show large differences for \Teff$\gtrsim 50$\,kK compared with the objects they had in 
common with the sample of \cite{Gianninas2011}. However, to distinguish whether a DA WD is hotter or cooler than \Teffw{45}, LTE models are probably good enough. 
By comparing pure He and H model atmosphere fluxes, we found that at \Teffw{45} DA WDs are about 0.1\,mag brighter than DO WDs. Accordingly, we set the magnitude 
limit for DA WDs to $g<17.4$ and for DO WDs to $g<17.5$. We found in total 117 DA WDs with \Teff$>$\,45\,kK, with 81 of them clean DA WDs (no subtypes, e.g. DAO, 
DAM, DAH). On the other hand we found 23 hot non-DA WDs in the SDSS DR7 spectroscopic sample with $g<17.5$, leading to DA/non-DA $=5.1$, with a lower limit of 
$3.5$ if we consider clean DA WDs alone.\\
Among the H-deficient objects before the wind limit that are included in the SDSS DR 10 spectroscopic sample, we find 18 PG\,1159 stars and 12 O(He) stars and DO 
WDs, which leads to a ratio of C-dominated to He-dominated objects of $1.5$. Including \logg$> 7.0$ objects alone, the ratio of C-dominated  to He-dominated 
objects is $0.75$. This suggests that DO WDs beyond the wind limit may be fed to a similar extent by PG\,1159  and O(He) stars.

\subsection{Conclusions}
\label{subsect:conclusions}

We have visually scanned a color-selected sample of white dwarf candidates in the SDSS DR10 and identified 22 new cool DO white dwarfs.
Effective temperatures, surface gravities, and C abundances (or at least upper limits) for 24 DO WDs were derived with non-LTE model atmospheres. 
Among the newly identified DO WDs, we found one more member of the so-called hot-wind DO WDs, which shows uhei absorption lines. One of the DO WDs is the most massive 
DO WD ever discovered with a mass of 1.07\Msol\ if it is an ONe-WD or 1.09\Msol\ if it is a CO-WD. Two of our objects are the coolest DO WDs ever discovered that 
still show  a considerable amount of C in the atmosphere ($C\,=0.001-0.01$). This strongly contradicts the diffusion 
calculations of \cite{UnglaubBues2000}. We suggest that -- similar to what has been proposed for the cooler DB stars -- a weak mass-loss is present in cool 
DO WDs. Furthermore, we presented the mass distribution of all hitherto analyzed DO WDs, PG\,1159, and O(He) stars. 
We found that the mass distribution of DO WDs beyond the wind limit strongly deviates from the mass distribution of the objects before the 
wind limit and explained this phenomenon with a scenario of different input channels. About 13\% of the DO WDs have masses below 0.5\,\Msol\ and might be successors of post-EHB 
He-sdO stars or low-mass PG\,1159 stars. The plateau around 0.6\,\Msol\ in the mass distribution of DO WDs beyond the wind limit agrees with the mass 
distribution of PG\,1159 stars, O(He) stars, and DO WDs before the wind limit. The two additional higher mass peaks might reflect a 
merger origin of some O(He) stars and DO WDs and/or the possibility that [WN] and [WC] type central stars are more massive than PG1159 and O(He) stars. 
The non-DA WD channel may be fed by about 13\% by post-EHB stars. PG\,1159 stars and O(He) stars may contribute to a similar extent to the non-DA WD channel.

\begin{acknowledgements}
NR is supported by the German Research Foundation (DFG, grant WE 1312/41-1),
TR by the German Aerospace Center (DLR, grant 05\,OR\,1401). 
The research leading to these results has received funding from the European Research Council under the European 
Union’s Seventh Framework Programme (FP/2007-2013) / ERC Grant Agreement n. 320964 (WDTracer). BTG was supported in 
part by the UK’s Science and Technology Facilities Council (ST/I001719/1).
Funding for SDSS-III has been provided by the Alfred P. Sloan Foundation, the Participating Institutions, the National 
Science Foundation, and the U.S. Department of Energy Office of Science. The SDSS-III web site is \url{http://www.sdss3.org/}.
SDSS-III is managed by the Astrophysical Research Consortium for the Participating Institutions of the SDSS-III Collaboration 
including the University of Arizona, the Brazilian Participation Group, Brookhaven National Laboratory, Carnegie Mellon University, 
University of Florida, the French Participation Group, the German Participation Group, Harvard University, the Instituto de Astrofisica 
de Canarias, the Michigan State/Notre Dame/JINA Participation Group, Johns Hopkins University, Lawrence Berkeley National Laboratory, 
Max Planck Institute for Astrophysics, Max Planck Institute for Extraterrestrial Physics, New Mexico State University, New York University, 
Ohio State University, Pennsylvania State University, University of Portsmouth, Princeton University, the Spanish Participation Group, 
University of Tokyo, University of Utah, Vanderbilt University, University of Virginia, University of Washington, and Yale University. 
This research has made use of the SIMBAD database, operated at CDS, Strasbourg, France. 
This research has made use of the VizieR catalogue access tool, CDS, Strasbourg, France.
\end{acknowledgements}

\bibliographystyle{aa}
\bibliography{aa} 

\begin{thebibliography}{75}
\expandafter\ifx\csname natexlab\endcsname\relax\def\natexlab#1{#1}\fi

\bibitem[{{Accad} {et~al.}(1971){Accad}, {Pekeris}, \& {Schiff}}]{Accad1971}
{Accad}, Y., {Pekeris}, C.~L., \& {Schiff}, B. 1971, \pra, 4, 516

\bibitem[{{Ahn} {et~al.}(2012){Ahn}, {Alexandroff}, {Allende Prieto},
  {Anderson}, {Anderton}, {Andrews}, {Aubourg}, {Bailey}, {Balbinot}, {Barnes},
  \& et~al.}]{Ahnetal2012}
{Ahn}, C.~P., {Alexandroff}, R., {Allende Prieto}, C., {et~al.} 2012, \apjs,
  203, 21

\bibitem[{{Althaus} {et~al.}(2005{\natexlab{a}}){Althaus},
  {Garc{\'{\i}}a-Berro}, {Isern}, \& {C{\'o}rsico}}]{Althaus2005b}
{Althaus}, L.~G., {Garc{\'{\i}}a-Berro}, E., {Isern}, J., \& {C{\'o}rsico},
  A.~H. 2005{\natexlab{a}}, \aap, 441, 689

\bibitem[{{Althaus} {et~al.}(2009){Althaus}, {Panei}, {Miller Bertolami},
  {Garc{\'{\i}}a-Berro}, {C{\'o}rsico}, {Romero}, {Kepler}, \&
  {Rohrmann}}]{althausetal2009}
{Althaus}, L.~G., {Panei}, J.~A., {Miller Bertolami}, M.~M., {et~al.} 2009,
  \apj, 704, 1605

\bibitem[{{Althaus} {et~al.}(2005{\natexlab{b}}){Althaus}, {Serenelli},
  {Panei}, {C{\'o}rsico}, {Garc{\'{\i}}a-Berro}, \&
  {Sc{\'o}ccola}}]{Althaus2005}
{Althaus}, L.~G., {Serenelli}, A.~M., {Panei}, J.~A., {et~al.}
  2005{\natexlab{b}}, \aap, 435, 631

\bibitem[{{Asplund} {et~al.}(2009){Asplund}, {Grevesse}, {Sauval}, \&
  {Scott}}]{asplundetal2009}
{Asplund}, M., {Grevesse}, N., {Sauval}, A.~J., \& {Scott}, P. 2009, \araa, 47,
  481

\bibitem[{{Barnard} {et~al.}(1969){Barnard}, {Cooper}, \&
  {Shamey}}]{Barnard1969}
{Barnard}, A.~J., {Cooper}, J., \& {Shamey}, L.~J. 1969, \aap, 1, 28

\bibitem[{{Barnard} {et~al.}(1974){Barnard}, {Cooper}, \&
  {Smith}}]{Barnard1974}
{Barnard}, A.~J., {Cooper}, J., \& {Smith}, E.~W. 1974, \jqsrt, 14, 1025

\bibitem[{{Bergeron} {et~al.}(2011){Bergeron}, {Wesemael}, {Dufour},
  {Beauchamp}, {Hunter}, {Saffer}, {Gianninas}, {Ruiz}, {Limoges}, {Dufour},
  {Fontaine}, \& {Liebert}}]{Bergeron2011}
{Bergeron}, P., {Wesemael}, F., {Dufour}, P., {et~al.} 2011, \apj, 737, 28

\bibitem[{{Bl\"ocker}(1995)}]{bloecker1995}
{Bl\"ocker}, T. 1995, \aap, 299, 755

\bibitem[{{Brassard} {et~al.}(2007){Brassard}, {Fontaine}, {Dufour}, \&
  {Bergeron}}]{Brassard2007}
{Brassard}, P., {Fontaine}, G., {Dufour}, P., \& {Bergeron}, P. 2007, in
  Astronomical Society of the Pacific Conference Series, Vol. 372, 15th
  European Workshop on White Dwarfs, ed. R.~{Napiwotzki} \& M.~R. {Burleigh},
  19

\bibitem[{{De Marco} {et~al.}(2014){De Marco}, {Long}, {George}, {Hillwig},
  {Kronberger}, {Howell}, {Reindl}, \& S.}]{DeMarco2014}
{De Marco}, O., {Long}, J., {George}, H.~J., {et~al.} 2014, \mnras, submitted

\bibitem[{{Desharnais} {et~al.}(2008){Desharnais}, {Wesemael}, {Chayer},
  {Kruk}, \& {Saffer}}]{Desharnais2008}
{Desharnais}, S., {Wesemael}, F., {Chayer}, P., {Kruk}, J.~W., \& {Saffer},
  R.~A. 2008, \apj, 672, 540

\bibitem[{{Dreizler}(1999)}]{Dreizler1999}
{Dreizler}, S. 1999, \aap, 352, 632

\bibitem[{{Dreizler} \& {Heber}(1998)}]{Dreizler1998}
{Dreizler}, S. \& {Heber}, U. 1998, \aap, 334, 618

\bibitem[{{Dreizler} {et~al.}(1995){Dreizler}, {Heber}, {Napiwotzki}, \&
  {Hagen}}]{Dreizleretal1995}
{Dreizler}, S., {Heber}, U., {Napiwotzki}, R., \& {Hagen}, H.~J. 1995, \aap,
  303, L53

\bibitem[{{Dreizler} \& {Werner}(1996)}]{dreizlerwerner1996}
{Dreizler}, S. \& {Werner}, K. 1996, \aap, 314, 217

\bibitem[{{Dufour} {et~al.}(2005){Dufour}, {Bergeron}, \&
  {Fontaine}}]{Dufour2005}
{Dufour}, P., {Bergeron}, P., \& {Fontaine}, G. 2005, \apj, 627, 404

\bibitem[{{Dufour} {et~al.}(2007){Dufour}, {Liebert}, {Fontaine}, \&
  {Behara}}]{Dufour2007}
{Dufour}, P., {Liebert}, J., {Fontaine}, G., \& {Behara}, N. 2007, \nat, 450,
  522

\bibitem[{{Dufour} {et~al.}(2002){Dufour}, {Wesemael}, \&
  {Bergeron}}]{Dufour2002}
{Dufour}, P., {Wesemael}, F., \& {Bergeron}, P. 2002, \apj, 575, 1025

\bibitem[{{Eisenstein} {et~al.}(2006){Eisenstein}, {Liebert}, {Harris},
  {Kleinman}, {Nitta}, {Silvestri}, {Anderson}, {Barentine}, {Brewington},
  {Brinkmann}, {Harvanek}, {Krzesi{\'n}ski}, {Neilsen}, {Long}, {Schneider}, \&
  {Snedden}}]{Eisenstein2006}
{Eisenstein}, D.~J., {Liebert}, J., {Harris}, H.~C., {et~al.} 2006, \apjs, 167,
  40

\bibitem[{{Engstrom} {et~al.}(1992){Engstrom}, {Bengtsson}, {Jupen}, \&
  {Westerlind}}]{Engstrom1992}
{Engstrom}, L., {Bengtsson}, P., {Jupen}, C., \& {Westerlind}, M. 1992, Journal
  of Physics B Atomic Molecular Physics, 25, 2459

\bibitem[{{Fontaine} \& {Brassard}(2005)}]{Fontaine2005}
{Fontaine}, G. \& {Brassard}, P. 2005, in Astronomical Society of the Pacific
  Conference Series, Vol. 334, 14th European Workshop on White Dwarfs, ed.
  D.~{Koester} \& S.~{Moehler}, 49

\bibitem[{{Frew} {et~al.}(2014){Frew}, {Boji{\v c}i{\'c}}, {Parker}, {Stupar},
  {Wachter}, {DePew}, {Danehkar}, {Fitzgerald}, \& {Douchin}}]{Frewetal2014}
{Frew}, D.~J., {Boji{\v c}i{\'c}}, I.~S., {Parker}, Q.~A., {et~al.} 2014,
  \mnras, 440, 1345

\bibitem[{{G{\"a}nsicke} {et~al.}(2010){G{\"a}nsicke}, {Koester}, {Girven},
  {Marsh}, \& {Steeghs}}]{Gaensicke2010}
{G{\"a}nsicke}, B.~T., {Koester}, D., {Girven}, J., {Marsh}, T.~R., \&
  {Steeghs}, D. 2010, Science, 327, 188

\bibitem[{{Gianninas} {et~al.}(2010){Gianninas}, {Bergeron}, {Dupuis}, \&
  {Ruiz}}]{Gianninas2010}
{Gianninas}, A., {Bergeron}, P., {Dupuis}, J., \& {Ruiz}, M.~T. 2010, \apj,
  720, 581

\bibitem[{{Gianninas} {et~al.}(2011){Gianninas}, {Bergeron}, \&
  {Ruiz}}]{Gianninas2011}
{Gianninas}, A., {Bergeron}, P., \& {Ruiz}, M.~T. 2011, \apj, 743, 138

\bibitem[{{Griem}(1974)}]{Griem1974}
{Griem}, H.~R. 1974, {Spectral line broadening by plasmas} (New York, Academic
  Press, Inc.~Pure and Applied Physics.~Volume 39, 1974.~421)

\bibitem[{{Hamann} {et~al.}(2003){Hamann}, {Pe{\~n}a}, {Gr{\"a}fener}, \&
  {Ruiz}}]{Hamann2003}
{Hamann}, W.-R., {Pe{\~n}a}, M., {Gr{\"a}fener}, G., \& {Ruiz}, M.~T. 2003,
  \aap, 409, 969

\bibitem[{{Heber} {et~al.}(1996){Heber}, {Dreizler}, \& {Hagen}}]{Heber1996}
{Heber}, U., {Dreizler}, S., \& {Hagen}, H.-J. 1996, \aap, 311, L17

\bibitem[{{Hirsch}(2009)}]{hirsch2009}
{Hirsch}, H.~A. 2009, PhD thesis, University Nuremberg

\bibitem[{{H{\"u}gelmeyer} {et~al.}(2006){H{\"u}gelmeyer}, {Dreizler},
  {Homeier}, {Krzesi{\'n}ski}, {Werner}, {Nitta}, \&
  {Kleinman}}]{huegelmeyeretal2006}
{H{\"u}gelmeyer}, S.~D., {Dreizler}, S., {Homeier}, D., {et~al.} 2006, \aap,
  454, 617

\bibitem[{{H{\"u}gelmeyer} {et~al.}(2005){H{\"u}gelmeyer}, {Dreizler},
  {Werner}, {Krzesi{\'n}ski}, {Nitta}, \& {Kleinman}}]{huegelmeyeretal2005}
{H{\"u}gelmeyer}, S.~D., {Dreizler}, S., {Werner}, K., {et~al.} 2005, \aap,
  442, 309

\bibitem[{{Iben} {et~al.}(1983){Iben}, {Kaler}, {Truran}, \&
  {Renzini}}]{Iben1983}
{Iben}, Jr., I., {Kaler}, J.~B., {Truran}, J.~W., \& {Renzini}, A. 1983, \apj,
  264, 605

\bibitem[{{Kleinman} {et~al.}(2013){Kleinman}, {Kepler}, {Koester}, {Pelisoli},
  {Pe{\c c}anha}, {Nitta}, {Costa}, {Krzesinski}, {Dufour}, {Lachapelle},
  {Bergeron}, {Yip}, {Harris}, {Eisenstein}, {Althaus}, \&
  {C{\'o}rsico}}]{Kleinman2013}
{Kleinman}, S.~J., {Kepler}, S.~O., {Koester}, D., {et~al.} 2013, \apjs, 204, 5

\bibitem[{{Koester} \& {Knist}(2006)}]{Koester2006}
{Koester}, D. \& {Knist}, S. 2006, \aap, 454, 951

\bibitem[{{Koester} {et~al.}(2014){Koester}, {Provencal}, \&
  {G{\"a}nsicke}}]{Koester2014}
{Koester}, D., {Provencal}, J., \& {G{\"a}nsicke}, B.~T. 2014, \aap, 568, A118

\bibitem[{{Lau} {et~al.}(2012){Lau}, {Gil-Pons}, {Doherty}, \&
  {Lattanzio}}]{Lauetal2012}
{Lau}, H.~H.~B., {Gil-Pons}, P., {Doherty}, C., \& {Lattanzio}, J. 2012, \aap,
  542, A1

\bibitem[{{Mahsereci}(2011)}]{mahsereci2011}
{Mahsereci}, M. 2011, Diploma thesis, University T\"ubingen, Germany

\bibitem[{{Miller Bertolami}(2014)}]{MillerBertolami2014b}
{Miller Bertolami}, M.~M. 2014, \aap, 562, A123

\bibitem[{{Miller Bertolami} {et~al.}(2014){Miller Bertolami}, {Melendez},
  {Althaus}, \& {Isern}}]{MillerBertolami2014a}
{Miller Bertolami}, M.~M., {Melendez}, B.~E., {Althaus}, L.~G., \& {Isern}, J.
  2014, ArXiv e-prints 1406.7712

\bibitem[{{Miszalski} {et~al.}(2012){Miszalski}, {Crowther}, {De Marco},
  {K{\"o}ppen}, {Moffat}, {Acker}, \& {Hillwig}}]{miszalskietal2012}
{Miszalski}, B., {Crowther}, P.~A., {De Marco}, O., {et~al.} 2012, \mnras, 423,
  934

\bibitem[{{Moore}(1970)}]{Moore1970}
{Moore}, C.~E. 1970, {Selected tables of atomic spectra} (NSRDS-NBS)

\bibitem[{{M\"uller-Ringat}(2013)}]{ringatPhD2013}
{M\"uller-Ringat}, E. 2013, Dissertation, University of T\"ubingen, Germany,
  http://tobias-lib.uni-tuebingen.de/volltexte/2013/6774/

\bibitem[{{Nagel} {et~al.}(2006){Nagel}, {Schuh}, {Kusterer}, {Stahn},
  {H{\"u}gelmeyer}, {Dreizler}, {G{\"a}nsicke}, \& {Schreiber}}]{Nageletal2006}
{Nagel}, T., {Schuh}, S., {Kusterer}, D.-J., {et~al.} 2006, \aap, 448, L25

\bibitem[{{Napiwotzki} \& {Sch{\"o}nberner}(1995)}]{NapiSchoen1995}
{Napiwotzki}, R. \& {Sch{\"o}nberner}, D. 1995, \aap, 301, 545

\bibitem[{{N{\'e}meth} {et~al.}(2012){N{\'e}meth}, {Kawka}, \&
  {Vennes}}]{Nemeth2012}
{N{\'e}meth}, P., {Kawka}, A., \& {Vennes}, S. 2012, \mnras, 427, 2180

\bibitem[{{Nousek} {et~al.}(1986){Nousek}, {Shipman}, {Holberg}, {Liebert},
  {Pravdo}, {White}, \& {Giommi}}]{Nousek1986}
{Nousek}, J.~A., {Shipman}, H.~L., {Holberg}, J.~B., {et~al.} 1986, \apj, 309,
  230

\bibitem[{{Pelletier} {et~al.}(1986){Pelletier}, {Fontaine}, {Wesemael},
  {Michaud}, \& {Wegner}}]{Pelletier1986}
{Pelletier}, C., {Fontaine}, G., {Wesemael}, F., {Michaud}, G., \& {Wegner}, G.
  1986, \apj, 307, 242

\bibitem[{{Petitclerc} {et~al.}(2005){Petitclerc}, {Wesemael}, {Kruk},
  {Chayer}, \& {Bill{\`e}res}}]{Petitclerc2005}
{Petitclerc}, N., {Wesemael}, F., {Kruk}, J.~W., {Chayer}, P., \&
  {Bill{\`e}res}, M. 2005, \apj, 624, 317

\bibitem[{{Provencal} {et~al.}(2000){Provencal}, {Shipman}, {Thejll}, \&
  {Vennes}}]{Provencal2000}
{Provencal}, J.~L., {Shipman}, H.~L., {Thejll}, P., \& {Vennes}, S. 2000, \apj,
  542, 1041

\bibitem[{{Provencal} {et~al.}(1996){Provencal}, {Shipman}, {Thejll}, {Vennes},
  \& {Bradley}}]{Provencal1996}
{Provencal}, J.~L., {Shipman}, H.~L., {Thejll}, P., {Vennes}, S., \& {Bradley},
  P.~A. 1996, \apj, 466, 1011

\bibitem[{{Rauch} \& {Deetjen}(2003)}]{rauchdeetjen2003}
{Rauch}, T. \& {Deetjen}, J.~L. 2003, in Astronomical Society of the Pacific
  Conference Series, Vol. 288, Stellar Atmosphere Modeling, ed. I.~{Hubeny},
  D.~{Mihalas}, \& K.~{Werner}, 103

\bibitem[{{Reindl} {et~al.}(2014){Reindl}, {Rauch}, {Werner}, {Kruk}, \&
  {Todt}}]{Reindletal2014}
{Reindl}, N., {Rauch}, T., {Werner}, K., {Kruk}, J.~W., \& {Todt}, H. 2014,
  \aap, 566, A116

\bibitem[{{Sch\"oning}(1993)}]{Schoening1993}
{Sch\"oning}, T. 1993, \aap, 267, 300

\bibitem[{{Sch\"oning} \& {Butler}(1989)}]{Schoening1989}
{Sch\"oning}, T. \& {Butler}, K. 1989, \aaps, 78, 51

\bibitem[{{Schuh} {et~al.}(2008){Schuh}, {Traulsen}, {Nagel}, {Reiff},
  {Homeier}, {Schwager}, {Kusterer}, {Lutz}, \& {Schreiber}}]{Schuh2008}
{Schuh}, S., {Traulsen}, I., {Nagel}, T., {et~al.} 2008, Astronomische
  Nachrichten, 329, 376

\bibitem[{{Sc{\'o}ccola} {et~al.}(2006){Sc{\'o}ccola}, {Althaus}, {Serenelli},
  {Rohrmann}, \& {C{\'o}rsico}}]{Scoccola2006}
{Sc{\'o}ccola}, C.~G., {Althaus}, L.~G., {Serenelli}, A.~M., {Rohrmann}, R.~D.,
  \& {C{\'o}rsico}, A.~H. 2006, \aap, 451, 147

\bibitem[{{Sinamyan}(2011)}]{Sinamyan2011}
{Sinamyan}, P.~K. 2011, Astrophysics, 54, 413

\bibitem[{{Sion}(2011)}]{Sion2011}
{Sion}, E.~M. 2011, {Hot White Dwarfs} (WILEY-VCH), 1

\bibitem[{{Todt} {et~al.}(2013){Todt}, {Kniazev}, {Gvaramadze}, {Hamann},
  {Buckley}, {Crause}, {Crawford}, {Gulbis}, {Hettlage}, {Hooper}, {Husser},
  {Kotze}, {Loaring}, {Nordsieck}, {O'Donoghue}, {Pickering}, {Potter},
  {Romero-Colmenero}, {Vaisanen}, {Williams}, \& {Wolf}}]{Todtetal2013}
{Todt}, H., {Kniazev}, A.~Y., {Gvaramadze}, V.~V., {et~al.} 2013, \mnras, 430,
  2302

\bibitem[{{Tremblay} {et~al.}(2013){Tremblay}, {Ludwig}, {Steffen}, \&
  {Freytag}}]{Tremblayetal2013}
{Tremblay}, P.-E., {Ludwig}, H.-G., {Steffen}, M., \& {Freytag}, B. 2013, \aap,
  559, A104

\bibitem[{{Unglaub} \& {Bues}(2000)}]{UnglaubBues2000}
{Unglaub}, K. \& {Bues}, I. 2000, \aap, 359, 1042

\bibitem[{{Wassermann} {et~al.}(2010){Wassermann}, {Werner}, {Rauch}, \&
  {Kruk}}]{wassermannetal2010}
{Wassermann}, D., {Werner}, K., {Rauch}, T., \& {Kruk}, J.~W. 2010, \aap, 524,
  A9

\bibitem[{{Werner}(1991)}]{Werner1991}
{Werner}, K. 1991, \aap, 251, 147

\bibitem[{{Werner} {et~al.}(2003){Werner}, {Deetjen}, {Dreizler}, {Nagel},
  {Rauch}, \& {Schuh}}]{werneretal2003}
{Werner}, K., {Deetjen}, J.~L., {Dreizler}, S., {et~al.} 2003, in Astronomical
  Society of the Pacific Conference Series, Vol. 288, Stellar Atmosphere
  Modeling, ed. I.~{Hubeny}, D.~{Mihalas}, \& K.~{Werner}, 31

\bibitem[{{Werner} {et~al.}(1995){Werner}, {Dreizler}, {Heber}, {Rauch},
  {Wisotzki}, \& {Hagen}}]{werneretal1995}
{Werner}, K., {Dreizler}, S., {Heber}, U., {et~al.} 1995, \aap, 293, L75

\bibitem[{{Werner} \& {Herwig}(2006)}]{wernerherwig2006}
{Werner}, K. \& {Herwig}, F. 2006, \pasp, 118, 183

\bibitem[{{Werner} \& {Rauch}(2014)}]{wernerrauch2014}
{Werner}, K. \& {Rauch}. 2014, \aap, 569, A99

\bibitem[{{Werner} {et~al.}(2004{\natexlab{a}}){Werner}, {Rauch}, {Barstow}, \&
  {Kruk}}]{Werner2004a}
{Werner}, K., {Rauch}, T., {Barstow}, M.~A., \& {Kruk}, J.~W.
  2004{\natexlab{a}}, \aap, 421, 1169

\bibitem[{{Werner} {et~al.}(2014){Werner}, {Rauch}, \&
  {Kepler}}]{werneretal2014}
{Werner}, K., {Rauch}, T., \& {Kepler}, S.~O. 2014, \aap, 564, A53

\bibitem[{{Werner} {et~al.}(2004{\natexlab{b}}){Werner}, {Rauch}, {Napiwotzki},
  {Christlieb}, {Reimers}, \& {Karl}}]{Werneretal2004}
{Werner}, K., {Rauch}, T., {Napiwotzki}, R., {et~al.} 2004{\natexlab{b}}, \aap,
  424, 657

\bibitem[{{Wood}(1994)}]{Wood1994}
{Wood}, M.~A. 1994, in Bulletin of the American Astronomical Society, Vol.~26,
  American Astronomical Society Meeting Abstracts, 1382

\bibitem[{{Zhang} \& {Jeffery}(2012{\natexlab{a}})}]{zhangetal2012b}
{Zhang}, X. \& {Jeffery}, C.~S. 2012{\natexlab{a}}, \mnras, 426, L81

\bibitem[{{Zhang} \& {Jeffery}(2012{\natexlab{b}})}]{zhangetal2012a}
{Zhang}, X. \& {Jeffery}, C.~S. 2012{\natexlab{b}}, \mnras, 419, 452

\end{thebibliography}

\end{document}